\pdfoutput=1
\documentclass[%
 aip,
 rsi,
amsmath,amssymb,
reprint,
floatfix
]{revtex4-1}

\usepackage{graphicx}
\usepackage{dcolumn}
\usepackage{bm}

\usepackage[utf8]{inputenc}
\usepackage{mathptmx}
\usepackage{braket}
\usepackage[english]{babel}
\usepackage{xcolor}
\colorlet{RED}{red}
\colorlet{BLUE}{blue}
\usepackage{dcolumn}
\usepackage{bm}
\usepackage[version=4]{mhchem} 
\usepackage{acronym}
\usepackage{adjustbox}
\usepackage{tikz}
\usepackage{microtype} 
\usetikzlibrary{calc,shapes.geometric,decorations.pathmorphing,patterns}

\definecolor{background-color}{gray}{0.98}
\usepackage[margin=2.3cm,bmargin=1cm,footnotesep=1cm]{geometry}
\usepackage{hyperref}
\hypersetup{
  colorlinks=true,
  linkcolor=blue,
  citecolor=blue,
  urlcolor=blue,
}
\usepackage{enumitem}
\begin{document}

\iftrue

This manuscript has been authored in part by UT-Battelle, LLC, under contract DE-AC05-00OR22725 with the US Department of Energy (DOE). The US government retains and the publisher, by accepting the article for publication, acknowledges that the US government retains a nonexclusive, paid-up, irrevocable, worldwide license to publish or reproduce the published form of this manuscript, or allow others to do so, for US government purposes. DOE will provide public access to these results of federally sponsored research in accordance with the DOE Public Access Plan (https://energy.gov/downloads/doe-public-access-plan).
\fi

\iftrue
This article may be downloaded for personal use only. Any other use
requires prior permission of the author and AIP Publishing. This
article appeared in volume 152, issue 18, page 184102 of the Journal of Chemical Physics and may be 
\fi
found at
\url{https://doi.org/10.1063/5.0004997}

\clearpage

\title[NWChem: Past, Present, and Future]{NWChem: Past, Present, and Future}



\author{E. Apr\`a}
\affiliation{Pacific Northwest National Laboratory, Richland, WA 99352, USA}

\author{E.~J. Bylaska}
\affiliation{Pacific Northwest National Laboratory, Richland, WA 99352, USA}

\author{W.~A. de Jong}
\affiliation{Lawrence Berkeley National Laboratory, Berkeley, CA 94720, USA}

\author{N. Govind}
\affiliation{Pacific Northwest National Laboratory, Richland, WA 99352, USA}

\author{K. Kowalski}
\email{karol.kowalski@pnnl.gov}
\affiliation{Pacific Northwest National Laboratory, Richland, WA 99352, USA}

\author{T.~P. Straatsma}
\affiliation{National Center for Computational Sciences, Oak Ridge National Laboratory, Oak Ridge, TN 37831, USA}

\author{M. Valiev}
\affiliation{Pacific Northwest National Laboratory, Richland, WA 99352, USA}

\author{H.~J.~J.  van Dam}
\affiliation{Brookhaven National Laboratory, Upton, NY 11973, USA}

\author{Y. Alexeev}
\affiliation{Argonne Leadership Computing Facility, Argonne National Laboratory, Argonne, IL 60439, USA}

\author{J. Anchell}
\affiliation{Intel Corporation, Santa Clara, CA 95054, USA}

% ...


\date{\today}

\begin{abstract}
Specialized computational chemistry packages have permanently reshaped the landscape of chemical and materials science by providing tools 
to support and guide experimental efforts and for the prediction of atomistic and electronic properties. 
In this regard, electronic structure packages have played a special role by using first-principle-driven methodologies to model complex chemical and materials processes. Over the last few decades, the rapid development of computing technologies and the tremendous increase in computational power have offered a unique chance to study complex 
transformations using sophisticated and predictive many-body techniques that describe correlated behavior of electrons in molecular and condensed phase systems at different levels of theory. In enabling these simulations,  novel parallel algorithms have been able to take advantage of computational resources to address the polynomial scaling of electronic structure methods. 
In this paper, we briefly review the NWChem computational chemistry suite, including its history, design principles, parallel tools, current capabilities, outreach and outlook.
\end{abstract}

\maketitle

\section{Introduction} 

The NorthWest Chemistry (NWChem) modeling software is a popular computational chemistry package that has been designed and developed to work efficiently on massively parallel processing  supercomputers\cite{bernholdt1995parallel,KENDALL2000260,VALIEV20101477}.  It contains an umbrella of modules that can be used to tackle most electronic structure theory calculations being carried out today. Since 2010, the code is  distributed as open-source under the terms of the Educational Community License version 2.0 (ECL 2.0).

Electronic structure theory provides a foundation for  our understanding of chemical transformations  and processes in complex chemical environments.
For this reason, accurate electronic structure formulations have already permeated several key areas of chemistry, biology, biochemistry, and materials sciences, where
they have become indispensable elements for building synergies between theoretical and experimental efforts and for predictions.
Over the last few decades, intense theoretical developments have resulted in a broad array of electronic structure methods and their implementations, designed to describe structures, interactions, chemical  reactivity, dynamics, thermodynamics, and spectral properties of molecular and material systems. 
The success  of these computational tools hinges upon several requirements regarding the accuracy of many-body models, 
reliable algorithms for dealing with processes at various spatial and temporal scales, and 
effective utilization of ever-growing computational resources. 
For instance,
the predictive power of computational chemistry requires sophisticated quantum mechanical approaches that systematically account for electronic correlation effects.
Therefore, the design of versatile electronic structure codes is a
major undertaking that requires close collaboration between experts
in theoretical and computational chemistry, applied mathematics, and computer science.

NWChem
~\cite{harrison2000nwchem,KENDALL2000260,straatsmanwchem,apra2005nwchem,bylaska2007nwchem,VALIEV20101477,nwchem.org},
like other widely used electronic structure programs,
was developed to fully realize the potential of computational modeling
to answer key scientific questions. It provides a wide range of
capabilities that can be deployed on supercomputing platforms to solve
two fundamental equations of quantum mechanics\cite{szabo2012modern,cramer2013essentials,jensen2017introduction} - time-independent and
time-dependent Schr\"odinger equations:
\begin{eqnarray} 
H|\Psi\rangle &=& E |\Psi\rangle \;,
\label{se1} \\
i\hbar \frac{\partial |\Psi\rangle}{\partial t} &=& H |\Psi\rangle \;,
\label{se2}
\end{eqnarray}
and a fundamental equation of Newtonian mechanics
\begin{equation}
m_i \mathbf{a}_i = \mathbf{F}_i \;,
\label{maf}
\end{equation}
where forces $\mathbf{F}_i$ include information about quantum effects.

Given the breadth of electronic structure theory, it does not come as
a surprise that equations (\ref{se1})-(\ref{se2}) can be solved using
various representations of quantum mechanics employing wavefunctions
($|\Psi\rangle$), electron densities ($\rho(\vec{r})$), or
self-energies ($\Sigma(\omega)$), which comprise the wide spectrum of NWChem's functionalities to compute the electronic
wavefunctions, densities, and associated properties of molecular and
periodic systems.  These functionalities include Hartree-Fock\cite{Fock1930a,slater_1930,Fock1930b,hartree_1935} self-consistent field
(SCF) and post-SCF correlated many-body approaches that build on the
SCF wavefunction to tackle static and dynamic correlation
effects. Among correlated approaches, NWChem offers second-order
M\o{}ller-Plesset perturbation theory; single- and multi-reference,
ground- and excited-state, and linear-response coupled-cluster (CC)
theories; multi-configuration self-consistent field (MCSCF); and
selected and full configuration interaction (CI) codes.  NWChem
provides extensive density functional theory\cite{hohenberg1964inhomogeneous,kohn1965self,ParrYang1994} (DFT) capabilities
with Gaussian and plane-wave basis set implementations. Within the Gaussian basis set framework, a broad range of DFT response properties, ground and excited-state molecular dynamics, linear-response (LR) and real-time (RT) time-dependent density functional theory (TDDFT) are available. The plane-wave DFT implementations offer the capability to 
run scalable \textit{ab initio} and Car-Parrinello molecular
dynamics~\cite{car1985unified}, and band-structure simulations. The plane-wave code supports
both norm-conserving~\cite{hamann1989generalized,troullier1991efficient,bachelet1982pseudopotentials} and projector augmented wave (PAW)~\cite{blochl1994projector}
pseudopotentials.

For all DFT methods outlined above, both analytical or numerical
gradients and Hessians are available for geometry optimization and
vibrational analysis.  Additionally, NWChem is capable of performing
classical molecular dynamics (MD) simulations using either AMBER or
CHARMM force fields. Through its modular design, the \textit{ab
initio} methods can be coupled with the classical MD to perform mixed
quantum mechanics and molecular mechanics simulations (QM/MM).
Various solvent models and relativistic approaches
are also available, with the spin-orbit contribution being only
supported at the Hartree-Fock (HF) and DFT levels of theory and
associated response properties. The NWChem functionality described is
only a subset of its full capabilities. We refer the reader to the
NWChem website\cite{nwchem.org} to learn about the full suite of
functionalities available to the user community.

Currently, NWChem is developed and maintained primarily by researchers
at the Department of Energy (DOE) Pacific Northwest National
Laboratory (PNNL), with help from researchers at other research institutions.
It has a broad user base, and it is being used across the
national laboratory system and throughout academia and industry around
the world.
In this paper, we provide a high-level overview of NWChem's
core capabilities, recent developments in electronic methods, and a
short discussion of ongoing and future efforts. We also illustrate the
strengths of NWChem stemming from the possibility of seamless
integration of methodologies at various scales and review scientific
results that would not otherwise be obtainable without using its
highly-scalable implementations of electronic structure methods.

\section{Brief History}

The NWChem project\cite{bernholdt1995parallel,anchell1998nwchem,KENDALL2000260,harrison2000nwchem,straatsmanwchem,apra2005nwchem,bylaska2007nwchem,VALIEV20101477,Straatsma2011}  started in 1992. It was originally designed and implemented as part of 
the construction project associated with the EMSL user facility at PNNL.
Therefore, the software project started around four years before the
EMSL computing center was up and running.
This raised challenges for the software developers working on the project, such as predicting the features of 
future hardware architectures and how to deliver high performing software, while maintaining programmer productivity.  Overcoming
these challenges led to a design effort that strove for flexibility and
extensibility, as well as high-level interfaces to functionality that hid some of the hardware issues from the chemistry software application
developer. Over the years, this design and implementation have successfully advanced multiple science agendas, and NWChem's extensive code base of more than 2 million lines provides high-performance, scalable software code with advanced scientific capabilities that are used throughout the molecular sciences community.

NWChem is an example of a co-design effort harnessing the expertise of researchers from multiple scientific disciplines to provide  users with computational chemistry tools that are scalable both in their ability to treat large scientific computational chemistry problems efficiently and in their use of computing resources
from high-performance parallel supercomputers to conventional workstation clusters.  In particular, NWChem has been designed to handle 
\begin{itemize} 
 \setlength{\itemsep}{0pt}%
 \setlength{\parskip}{0pt}%
\item biomolecules, nanostructures, interfaces, and solid-state,
\item chemical processes in complex environments, 
\item hybrid quantum/classical simulations, 
\item ground and excited-states and non-linear optical properties,
\item simulations of UV-Vis, photo-electron, X-ray spectroscopies,
\item Gaussian basis functions or plane-waves,
\item ab-initio molecular dynamics on the ground and excited states,
\item relativistic effects.
\end{itemize}

The scalability of NWChem has provided a computational platform to deliver new scientific results that would be unobtainable if parallel computational platforms were not used. 
For example, NWChem's implementation of a non-orthogonally spin adapted CCSD(T) method has been demonstrated to scale to 210,000 processors available at the Oak Ridge National Laboratory's (ORNL) Leadership Computing Facilities,\cite{apra2009liquid,tipparaju2010enabling,kowalski2011scalable} whereas the plane-wave DFT code has been able to utilize close to 100,000 processor cores on NERSC's Cray-XE6 supercomputer.\cite{bylaska2011large}  
Although implemented only for the perturbative part of coupled-cluster with singles and doubles (CCSD)
\cite{purvis82_1910} and triples correction (CCSD(T)),\cite{Raghavachari} 
NWChem was one of the first computational chemistry codes to have been ported to utilize graphics processing units (GPUs).\cite{ma2011gpu}
Several parts of the code have also been rewritten to take advantage of the Intel Xeon Phi family of processors - good scalability and performance have been demonstrated for the \textit{ab initio} molecular dynamics plane-wave DFT code on the most recent Knights Landing version of the processor.\cite{bylaskaintel17,bylaska2017transitioning} The non-iterative triples part of the CCSD(T) method has been demonstrated to scale to 55,200 Intel Phi threads and 62,560 cores through concurrent utilization of CPU and Intel Xeon Phi Knights Corner accelerators.\cite{apra2014efficient}

\section{Design Principles}
NWChem has a five-tiered modular architecture.
The first tier is the {\em Generic Task Interface}. This interface
(an abstract programming interface, not a user interface)
serves as the
mechanism that transfers control to the different modules in the second tier,
which consists of the {\em Molecular Calculation Modules}.
The molecular calculation modules are the high-level programming
modules that accomplish computational tasks, performing particular operations
using the specified theories defined by the user in the input file.  These independent modules
of NWChem share data only through a disk-resident database,
which allows modules to share data or to share access to files containing
data.
The third tier consists of the {\em Molecular
Modeling Tools}.  These routines provide basic chemical functionality such as symmetry,
basis sets, grids, geometry, and integrals.
The fourth tier is
the {\em Software
Development Toolkit}, which is the basic foundation of the code.
The fifth tier provides the {\em Utility Functions} needed by nearly all modules
in the code.  These include such functionality as input processing, output processing,
and timing.
The {\em Generic Task Interface} controls the execution of NWChem.  The flow 
of control proceeds in the following steps:

\begin{enumerate}
\setlength{\itemsep}{0pt}%
\setlength{\parskip}{0pt}
\item Identify and open the input file.
\item Complete the initialization of the parallel environment.
\item Process start-up directives.
\item Summarize start-up information and write it to the output file.
\item Open the run-time database.
\item Process the input sequentially (ignoring start-up directives), 
including the first \verb+task+ directive.
\item Execute the task.
\item Repeat steps 6 and 7 until reaching the end of the input file or
encountering a fatal error condition.
\end{enumerate}

The input parser processes the user's input file and translates the
information into a form meaningful to the main program and the driver routines
for specific tasks.

As mentioned in step 5 of the task flow control, NWChem makes use of a
run-time database to store the main computational parameters.
This is in the same spirit of check-pointing features available in other quantum chemistry codes.
The information stored in the run-time database can be used at a later time in order
to restart a calculation. Restart capabilities are available for most modules. For example,
SCF generated files (run-time database and molecular orbitals) can be
used either to continue a geometry optimization or to compute molecular
properties. The important second and fourth tiers are discussed as part of the subsequent sections. 

%
%
%
\section{Parallel Tools}
\label{section:parallel}

The design and early development of Global Arrays\cite{ga1995,ga1995,ga1996,ga2006,gamanual2012} (GA) toolkit
occurred in the same period when the NWChem project started.
The GA toolkit, which is the central component of the {\em Software Development Toolkit}, was adopted by the NWChem developers as the main approach for the parallelization of the dense matrices
present in quantum chemistry methods that make use of local basis functions.
In current computer science parlance, Global Arrays can be viewed as
a Partitioned Global Address Space (PGAS) model that provides a high level
of abstraction for the programmer to the dense distributed arrays.
In contrast to message passing constructs such as MPI, where the developer
has to worry about coordinating send and receive operations, the use of
Global Arrays in NWChem requires so-called single-sided functions
(e.g. put, get, accumulate) to manipulate data structures  in a single operation.
The choice of distribution model for sharing a given global array among the memory available to the processes in use plays a crucial role in efficient parallelization at large scale.

The GA toolkit has been ported to a variety of parallel computer architectures. The porting process has focused in the past in optimizing the ARMCI\cite{armci2006} library.
The Aggregate Remote Memory Copy (ARMCI) library optimizes performance by fully exploiting network characteristics such as latency, bandwidth, and packet injection rate through the use of low-level network protocols (e.g. Infiniband Verbs). More recent porting options make use either of
ComEx\cite{comex2014} or
of the ARMCI-MPI  \cite{dinan2012supporting} communication runtimes.
Both ComEx and ARMCI-MPI make use of MPI libraries, instead of low-level network protocols,
albeit with different approaches.


%
%
%
%
%
%
\section{Main Methodologies}

In this section, we describe the key methods that comprise the {\em Molecular Calculation Modules}. We first describe the Gaussian basis HF and DFT
implementations for molecular systems. This is followed by the
post-SCF wavefunction-based perturbative (MP2), multi-configuration SCF, 
and high-accuracy (coupled-cluster theory) approaches for molecules,
including the tensor contraction engine (TCE). Molecular response
properties and relativistic approaches are then described. The
plane-wave based DFT implementation for Car-Parrinello molecular
dynamics and periodic condensed phase systems is described next,
followed by classical molecular dynamics and hybrid methods.

\subsection{Hartree-Fock}
The NWChem SCF module computes closed-shell
restricted Hartree-Fock (RHF) wavefunctions, restricted high-spin
open-shell Hartree-Fock (ROHF) wavefunctions, and spin-unrestricted Hartree-Fock (UHF) wavefunctions. The Hartree-Fock equations are solved using a conjugate gradient method with an orbital Hessian based preconditioner\cite{doi:10.1002/jcc.540161010}.

The most expensive part to compute in the SCF code is the two-electron contribution to the matrix element of the Fock operator (resulting
from the sum of Coulomb and Exchange operators).
To compute these matrix elements, NWChem developers have implemented parallel algorithms
using either a distributed data approach\cite{doi:10.1002/(SICI)1096-987X(19960115)17:1<109::AID-JCC9>3.0.CO;2-V}
(where the Fock matrix is distributed among the aggregate memory of the processes involved in the calculation) or a
replicated data approach (where an entire  copy of the Fock matrix is stored in memory of each process).

Several options are available for the initial guess of the SCF calculations.
The default choice uses the eigenvectors of a Fock-like matrix formed from a superposition of the atomic densities.
Other options include the  use of eigenvectors of the bare-nucleus Hamiltonian or the one-electron Hamiltonian,
the projections of  molecular orbital from a smaller basis to a larger one, or molecular orbitals  formed by superimposing the orbitals of fragments of the molecule being studied. Symmetry can be used to speed up the Fock matrix construction via the petite-list algorithm.
Molecular orbitals are symmetry adapted as well in NWChem. The resolution of the identity (RI) four-center, two-electron integral
approximation has also been implemented.\cite{Kendall1997}

In order to avoid full matrix diagonalization, 
the SCF program uses a preconditioned conjugate gradient (PCG) method
that is unconditionally convergent. Basically, a search direction is
generated by multiplying the orbital gradient (the derivative of the
energy with respect to the orbital rotations) by an approximation to
the inverse of the level-shifted orbital Hessian. In the initial
iterations, an inexpensive
one-electron approximation to the inverse orbital Hessian is
used. Closer to convergence, the full orbital Hessian is used, which
should provide quadratic convergence. For both the full or
one-electron orbital Hessians, the inverse-Hessian matrix-vector
product is formed iteratively. Subsequently, an approximate line
search is performed along the new search direction. 

Both all-electron basis sets and effective core potentials (ECPs) can be used.
Effective core potentials are a useful means of replacing the
core electrons in a calculation with an effective potential, thereby
eliminating the need for the core basis functions, which usually
require a large set of Gaussians to describe them. In addition to
replacing the core, they may be used to represent relativistic
effects, which will be discussed later.

%
\subsection{Density Functional Theory}
The NWChem DFT module for molecular systems uses a Gaussian basis set
to compute closed- and open-shell densities and Kohn-Sham orbitals in
the local density approximation (LDA), generalized gradient
approximation (GGA), $\tau$-dependent and Laplacian-dependent
meta-generalized gradient approximation (metaGGA), any combination of
local and non-local approximations (including exact exchange and
range-separated exchange), and asymptotically corrected
exchange-correlation potentials. NWChem contains energy-gradient
implementations of most exchange-correlation functionals available in
the literature, including a flexible framework to combine different
functionals. However, second derivatives are not supported for
meta-functionals and third derivatives are supported only for a
selected set of functionals. For a detailed description, we refer the
reader to the online documentation\cite{xcsummary}.


The DFT module reuses elements of the Gaussian basis SCF module for the evaluation of the
Hartree-Fock exchange and of the Coulomb matrices
by using 4-index 2-electron electron repulsion integrals;
the formal scaling of the DFT computation can be reduced by choosing to use auxiliary Gaussian
basis sets to fit the charge density\cite{dunlap79}
and use 3-index 2-electron integrals, instead. 

The DFT module supports both a distributed data approach and a mirrored arrays\cite{mirroredga} approach for the evaluation of
the exchange-correlation potential and energy. The mirrored arrays option, used by default, allows the calculation to hide network communication overhead by replicating the data between processes belonging to the same network node.

In analogy with what is available in the SCF module, the DFT module can perform restricted closed-shell,
unrestricted open-shell, and restricted open-shell calculations. However, in contrast to the SCF module that uses PCG to solve the SCF equation,
the DFT module implements diagonalization with parallel eigensolvers.
\cite{peigs1993,peigs1995,peigs1997,scalapack,elpa1,elpa2}
DIIS (direct inversion in the iterative subspace or direct inversion of the iterative subspace)\cite{pulay80},
level-shifting\cite{saunders_ls1,saunders_ls2} and density matrix damping can be used to
accelerate the convergence of the iterative SCF process. Another technique that can be used to help SCF convergence makes use of electronic smearing of the molecular orbital occupations, by using a gaussian broadening function following the prescription of Warren and Dunlap\cite{WarrenDunlap1996}. Additionally, calculations with fractional numbers of electrons can be performed to analyze the behavior of exchange-correlation functionals and their impact on molecular excited states and response properties.\cite{refaely2012,stein2012,srebro2012,moore2012,autschbach2014,sun2013,sun2014,moore2015}

The Perdew and Zunger\cite{perdew1981self}
method to remove the self-interaction contained in
many exchange-correlation functionals has been implemented\cite{garza2000} within the
Optimized Effective Potential (OEP) method\cite{sharp1953variational,
talman1976optimized} and
 within the Krieger-Li-Iafrate (KLI) approximation.\cite{krieger1992construction,li1993self}

The asymptotic region of the exchange-correlation
potential can be modified by the van-Leeuwen-Baerends exchange-correlation potential
that has the correct $-\frac{1}{r}$ asymptotic behavior. The total energy is then computed using the definition of the exchange-correlation
functional. This scheme is known to tend to over-correct the deficiency
of most uncorrected exchange-correlation potentials\cite{cs00,hiratazhan} and can improve TDDFT-based excitation calculations, but it is not variational. A variationally consistent approach to address this issue is via range-separated exchange-correlation functionals and the recently developed nearly correct asymptotic potential or NCAP\cite{carmona2018-ncap}, which are implemented in NWChem.

To describe dispersion interactions, both the exchange-hole dipole moment dispersion model (XDM)\cite{xdm}
and Grimme's DFT-D3  dispersion correction
(both zero-damped and BJ-damped variants)
for DFT functionals \cite{grimme2010,grimme2011} are available.
In many cases, one can obtain reasonably accurate non-covalent interaction energies at van der Waals distances with
meta-functionals in NWChem even without adding extra dispersion terms.\cite{zhao_2011}

Numerical integration is necessary for the evaluation of the
exchange-correlation contribution to the density functional when Gaussian basis functions are used.
The three-dimensional molecular integration problem is reduced to a sum of atomic integrations
by using the approach first proposed by Becke\cite{becke1988multicenter}.
NWChem implements a modification of the Stratmann  algorithm \cite{stratmann1996},
where the polynomial partition function $w_A(r)$ is replaced by a modified error function erf$n$ 
(where $n$ can be 1 or 2).

\begin{eqnarray*}
 w_A(r) & = & \prod_{B\neq A}\frac{1}{2} \left[1 \ - erf(\mu^\prime_{AB})\right] \\
 \mu^\prime_{AB} & = & \frac{1}{\alpha} \ \frac{\mu_{AB}}{(1-\mu_{AB}^2)^n}\\
 \mu_{AB} & = & \frac{{\mathbf r}_A - {\mathbf r}_B}{\left|{\mathbf r}_A - {\mathbf r}_B \right|}
\end{eqnarray*}

The default quadrature used for the atomic centered numerical integration is an
Euler-MacLaurin scheme for the radial components (with a modified
Mura-Knowles\cite{muraknowles1996} transformation) and a Lebedev\cite{lebedev1999quadrature} scheme for the angular
components.

On top of the petite-list symmetry algorithm used in the same fashion as in the SCF module,
the evaluation of the exchange-correlation kernel incurs additional time savings when the molecular symmetry is
a subset of the $O_h$ point group, exploiting the octahedral symmetry of the Lebedev angular grid.


NWChem also has an implementation of a variational
treatment of the one-electron spin-orbit operator within the DFT framework. Calculations can be performed either with an all-electron relativistic approach (for example, ZORA) or with an ECP and a matching spin-orbit (SO) potential. 

Other capabilities built on the DFT module include the electron transfer (ET)\cite{faradzel1990,rosso2003}, constrained DFT (CDFT) \cite{cdft1,cdft2,cdft3}, and frozen density embedding (FDE)\cite{wesolowski1993,wesolowski2008,doi:10.1021/acs.jctc.8b01036} modules, respectively.

%
\subsubsection{Time-Dependent Density Functional Theory}

\paragraph{Linear-Response Time-Dependent Density Functional Theory:}

NWChem supports a spectrum of single excitation theories for vertical excitation energy calculations, namely, configuration interaction singles (CIS)\cite{excited_1}, time-dependent Hartree-Fock (TDHF or also known as random-phase approximation RPA), time-dependent density functional theory (TDDFT)\cite{excited_2a,excited_2b,excited_2c}, and Tamm-Dancoff approximation\cite{excited_3} to TDDFT. These methods are implemented in a single framework that invokes Davidson's trial vector algorithm (or its modification for a non-Hermitian eigenvalue problem). An efficient special symmetric Lanczos algorithm and kernel polynomial method has also been implemented.\cite{jirkachao2017}

In addition to valence vertical excitation energies, core-level
excitations\cite{lopata2012} and emission
spectra\cite{zhang2015,zhang2019} can also be computed. Analytical
first derivatives of vertical excitation energies with a selected set
of exchange-correlation functionals can also be
computed,\cite{silverstein2013} which allows excited-state
optimizations and dynamics. Origin-independent optical rotation and
rotatory strength tensors can also be calculated with the LR-TDDFT
module within the gauge including atomic orbital (GIAO) basis
formulation.\cite{srebro2011,autschbach-4,moore2012,zhang2017}
Extensions to compute excited-state couplings are currently underway
and will be available in a future release.\\

\paragraph{Real-Time Time-Dependent Density Functional Theory:}

Real-time time-dependent density functional theory (RT-TDDFT) is a DFT-based approach to electronic excited states based on integrating the time-dependent Kohn-Sham (TDKS) equations in time. The theoretical underpinnings, strengths, and limitations are similar to traditional linear-response (LR) TDDFT methods, but instead of a frequency domain solution to the TDKS equations, RT-TDDFT yields a full time-resolved, potentially non-linear solution. Real-time simulations can be used to
compute not only spectroscopic properties (e.g., ground and excited-state absorption spectra, polarizabilities, etc.)\cite{lopata2011,lopata2012,tussupbayev2015,fischer2015,bowman2017}, but also the time and space-resolved electronic response to arbitrary external stimuli (e.g., electron charge dynamics after laser excitation)\cite{lopata2011,bruner2017} and non-linear spectroscopies.\cite{cho2018phase,bruner2019} RT-TDDFT has the potential to be efficient for computing spectra in systems with a high density of states\cite{wang2013optical} as, in principle, an entire absorption spectrum can be computed from only one dynamics simulation.

This functionality is developed on the Gaussian basis set DFT module for both restricted and unrestricted calculations and can be run with essentially any combination of basis set and exchange-correlation functional in NWChem. A number of time propagation algorithms have been implemented\cite{castro2004} within this module, with the default being the Magnus propagator.\cite{magnus1954} Unlike LR-TDDFT, which requires second derivatives, RT-TDDFT can be used with all the functionals since only first derivatives are needed for the propagation. The current RT-TDDFT implementation assumes frozen nuclei and no dissipation.

\subsubsection{Ab Initio Molecular Dynamics}
This module leverages the Gaussian basis set methods to allow for seamless molecular dynamics of molecular systems. The nuclei are treated as classical point particles and their motion is integrated via the velocity Verlet algorithm.\cite{Verlet67_98,Wilson82_637} In addition to being able to perform simulations in the microcanonical ensemble, we have implemented several thermostats to control the kinetic energy of the nuclei. These include the stochastic velocity rescaling approach of Bussi, Donadio, and Parrinello\cite{Parrinello07_014101}, Langevin dynamics according to the implementation of Bussi and Parrinello\cite{Parrinello07_056707}, the Berendsen thermostat\cite{Berendsen1984}, and simple velocity rescaling.

The potential energy surface upon which the nuclei move can be provided by any level of theory implemented within NWChem, including DFT, TDDFT, MP2, and the correlated wavefunction methods in the TCE module. If analytical gradients are implemented for the specified method, these are automatically used. Numerical gradients will be used in the event that analytical gradients are not available at the requested level of theory. This module has been used to demonstrate how the molecular dynamics based determination of vibrational properties can complement those determined through normal mode analysis, therefore allowing to achieve a deeper understanding of complex dynamics and to help interpret complex experimental signatures.\cite{Govind16_1429} Extensions to include non-adiabatic dynamics have been implemented in a development version and will be available in a future release.

\subsection{Wavefunction Formulations}
The wavefunction-based methods play a special role in all electronic structure packages. 
Their strengths originate in the 
possibility of introducing,   using either various orders of perturbation theory or  equivalently through the linked cluster theorem 
(for example, see Refs. \onlinecite{lindgren2012atomic} and \onlinecite{shavitt2009many})
various ranks of excitations, a systematic hierarchy of electron correlation effects. 
NWChem offers implementations of several correlated wavefunction approaches including 
many-body perturbation theory approaches and coupled-cluster methods.
 
\subsubsection{Perturbative Formulations} 
\paragraph{MP2:}
Three algorithms are available in NWChem to compute the M\o{}ller-Plesset (or
many-body) perturbation theory second-order correction\cite{PhysRev.46.618}
to the Hartree-Fock energy (MP2). They vary in capability, the size of the
system that can be treated and use of other approximations
\begin{itemize}
\setlength{\itemsep}{0pt}%
\setlength{\parskip}{0pt}
\item Semi-direct MP2 is recommended for most large applications
  on parallel computers with significant disk I/O capability. Partially
  transformed integrals are stored on disk, multi-passing as
  necessary. RHF and UHF references may be treated including
  computation of analytic derivatives. The initial semi-direct code was later modified
  to use aggregate memory instead of disk to store intermediate, therefore not requiring any I/O operation.

\item Fully-direct\cite{MP2Wong1996} MP2. This is of utility if only
  limited I/O resources are available (up to about 2800
  functions). Only RHF references and energies are available.

\item The resolution of the identity (RI) approximation MP2 (RI-MP2)\cite{BernholdtRIMP2}
  uses the RI approximation and is, therefore, only exact in the
  limit of a complete fitting basis. However, with some care, high
  accuracy may be obtained with relatively modest fitting basis
  sets. An RI-MP2 calculation can cost over 40 times less than the
  corresponding exact MP2 calculation. RHF and UHF references with
  only energies are available. 
\end{itemize}

\iftrue
\subsubsection{Multi-configurational Self-Consistent Field (MCSCF)}
 A large-scale parallel multi-configurational self-consistent field
 (MCSCF) method has been developed in NWChem by integration of the
 serial LUCIA program of Olsen\cite{olsen_1990,vogiatzis_2017}.
 The generalized active space approach is used to partition large
 configuration interaction (CI) vectors and generate a sufficient
 number of nearly equal batches for parallel distribution. This
 implementation allows the execution of complete active space
 self-consistent field (CASSCF) calculations with non-conventional
 active spaces. An unprecedented CI step for an expansion composed of
 almost one trillion Slater determinants has been reported\cite{vogiatzis_2017}.

\fi

\subsubsection{Coupled-Cluster Theory}

The coupled-cluster module of NWChem contains two classes of implementations (a) parallel implementation of the CCSD(T) formalism 
\cite{Raghavachari}
for closed-shell systems, and (b) a wide array of CC formalisms for arbitrary reference functions. The latter class of implementations automatically generated by Tensor Contraction Engine 
\cite{hirata2003tensor,1386652}
is an example of a successful co-design effort. 

\paragraph{Closed-Shell CCSD(T):} 
\label{ccsd}
The coupled-cluster method was introduced to chemistry by \v{C}\'{i}\v{z}ek \cite{cizek66_4256}
(see also Ref. \onlinecite{cizek}), 
and is a post-Hartree-Fock electron correlation method. Development of the canonical coupled-cluster code in NWChem commenced in 1995 under a collaboration with Cray Inc to develop a massively parallel coupled-cluster program designed to run on a Cray T3E.
Full details of the implementation are given in Kobayashi and Rendell\cite{Kobayashi}.

The coupled-cluster wavefunction is written as an exponential of excitation operators acting on the Hartree-Fock reference:
\begin{eqnarray}
\vert \Psi_{\rm CC} \rangle = e^T \vert \Phi \rangle
\label{ccsd_eq1}
\end{eqnarray}
where $T= T_1 + T_2 + . . .$ is a cluster operator represented as a sum of its many-body components, i.e., singles $T_1$, doubles $T_2$, etc. and $|\Phi\rangle$ 
is the so-called reference function (usually chosen as a reference determinant). 
In practical applications
the above sum is  truncated at some excitation rank. For example, the CCSD method \cite{purvis82_1910} 
is defined by including singles and doubles, i.e., $T\simeq T_1+T_2$. 
Introducing the exponential ansatz (\ref{ccsd_eq1}) into the Schr{\"o}dinger equation, premultiplying both sides by $e^{-T}$, using the Hausdorff formula,
and projecting onto the subspace of excitation functions, gives a set of coupled non-linear equations that are solved iteratively to yield the
coupled-cluster energy and amplitudes. For example, for the CCSD formulation one obtains
\begin{eqnarray}
\langle \Phi \vert (H_Ne^{T_1+T_2})_C \vert  
\Phi\rangle &=& \Delta E_{\rm CCSD}\label{ccsd_eq2} \\
\langle \Phi^{a}_{i} \vert (H_Ne^{T_1+T_2})_C |\Phi\rangle &=&  0 \;, \label{ccsd_eq3}\\
\langle \Phi^{ab}_{ij} \vert (H_Ne^{T_1+T_2})_C |\Phi\rangle &=&  0 \;,
\label{ccsd_eq4}
\end{eqnarray}
where $H_N$ is the electronic Hamiltonian in normal product form ($H_N=H-\langle\Phi|H|\Phi\rangle$),  subscript $C$ represents a connected part of a given operator expression, and $\Delta E_{\rm CCSD}$ is CCSD correlation energy. 
The closed-shell CCSD implementation employs  the optimized form of the CC equations discussed by  Scuseria {\it et al.} \cite{Scuseria} as was programmed in the TITAN program \cite{TITAN}.
The nature of the Cray T3E hardware required significant rewriting of earlier coupled-cluster algorithms to take into account the limited memory available per core (8 MW) and the prohibitive penalty of I/O operations. 
Of the various four indexed quantities, those with four occupied indices were replicated in local memory (i.e. the memory associated with a single core), and those with one or two virtual indices were distributed across the global memory of the machine (i.e. the sum of
the memory of all the processors), and accessed in computational batches. The terms involving integrals with three and four virtual orbital indices still proved too costly for the available memory and to circumvent this problem, these terms were evaluated in a "direct" fashion.
This structure distinguishes NWChem from most other coupled-cluster programs. Thus, to make effective use of the available memory, as much as possible should be allocated, by using global arrays, with the bare minimum for the arrays replicated in local memory.

The canonical CCSD implementation in NWChem also contains the perturbative triples correction, denoted (T), of
Raghavachari {\it et al.} \cite{Raghavachari}. This correction is an estimate from M{\o}ller-Plesset perturbation theory \cite{PhysRev.46.618} and evaluates the triples contribution to MP4 using the optimized cluster amplitudes
at the end of a CCSD calculation. The CCSD(T) method is commonly referred to as the gold standard for {\it ab initio} electronic structure theory calculations. Its computational cost scales as $n^7$, making it considerably more expensive than a CCSD calculation. However, the triples are non-iterative and only require two-electron integrals with at most three virtual orbital indices, hence avoiding the previous memory and I/O issues and so the correction was easily adapted from the "aijkbc
algorithm" of an earlier work by Rendell {\it et al} \cite{Rendell}. 

In recent years, a great deal of effort was invested to enhance the performance  of the iterative and non-iterative parts of the CCSD(T) workflow. Performance tuning of the iterative part resulted in scaling the code up to 223,200 processors of the ORNL Jaguar computer.\cite{apra2009liquid,h2o17_2010}
Significant speedups for the CCSD iterative part were achieved by introducing efficient optimization techniques to alleviate the communication bottlenecks caused by a copious amount of communication requests introduced by a large class of low-dimensionality tensor contractions.
This optimization provided a significant  2- to 5-fold performance increase 
in the CCSD iteration time depending on the problem size and available memory, and improved the CCSD scaling to 20,000 nodes of the NCSA Blue Waters supercomputer \cite{anisimov2014optimization}. \\

\paragraph{Tensor Contraction Engine and High-Accuracy Formulations:}
\label{section:tce}
NWChem implements a large number of high-rank electron-correlation methods for the ground, excited, and electron-detached/attached states as well as for molecular properties. The underlying ansatzes span configuration interaction (CI), coupled-cluster (CC), many-body perturbation theories (MBPT), and various combinations thereof. A distinguishing feature of these implementations is their uniquely forward-looking development strategy. These parallel-executable codes, as well as their formulations and algorithms, were computer-generated by the symbolic algebra program\cite{hirata2006symbolic} called Tensor Contraction Engine (TCE).\cite{hirata2003tensor} 
TCE was one of the first  attempts to provide a scalable tensor library for parallel implementations of many-body methods, which  extends the ideas of automatic CC code generation introduced by Janssen and Schaefer,\cite{janssen1991} Li and Paldus,\cite{li1994automation} and Nooijen and co-workers.\cite{nooijen2001towards,nooijen2002state}

The merits of such a symbolic system are many: (1) It expedites otherwise time-consuming and error-prone derivation and programming processes; (2) It facilitates parallelization and other laborious optimizations of the synthesized programs; (3) It enhances the portability, maintainability,  extensibility, and thus the lifespan of the whole program module; (4) It enables new or higher-ranked methods to be implemented and tested rapidly which are practically impossible to write manually. TCE is, therefore, one of the earliest examples\cite{janssen1991} of an expert system that lifts the burden of derivation/programming labor so that computational chemists can focus on imagining new ansatz---a development paradigm  embraced quickly by other chemistry software developers.\cite{parkhill2009,deumens2011software,macleod2015}

The working equations of an {\it ab initio} electron-correlation method are written with sums-of-products of matrices, whose elements are integrals of operators in the Slater determinants. For many methods, the matrices have the general form:\cite{hirata2004higherex}
\begin{eqnarray}
\langle \Phi_i | \hat{L}_j^\dagger \,\hat{H}\, \exp(\hat{T}_k)\,\hat{R}_l |\Phi_m \rangle_{\text{C/L}},
\label{TCE_eq1}
\end{eqnarray}
where $\Phi_i$ is the whole set of the $i$-electron excited (or electron-detached/attached) Slater determinants, $\hat{H}$ is the Hamiltonian operator, $\hat{T}_k$ is a $k$-electron excitation operator, $\hat{R}_l$ is an $l$-electron excitation (or electron detachment/attachment) operator, and $\hat{L}_j^\dagger$ is a $j$-electron de-excitation (or electron detachment/attachment) operator. Subscript `C/L' means that the operators can be required to be connected and/or linked diagrammatically. For example, the so-called $T_2$-amplitude equation of coupled-cluster singles and doubles (CCSD) is written as
\begin{eqnarray}
0 = \langle \Phi_2 | \hat{H}\, \exp(\hat{T}_1 + \hat{T}_2) |\Phi_0 \rangle_{\text{C}}.
\label{TCE_eq2}
\end{eqnarray}

\begin{figure}[ht]
\centering
\includegraphics[scale=0.40]{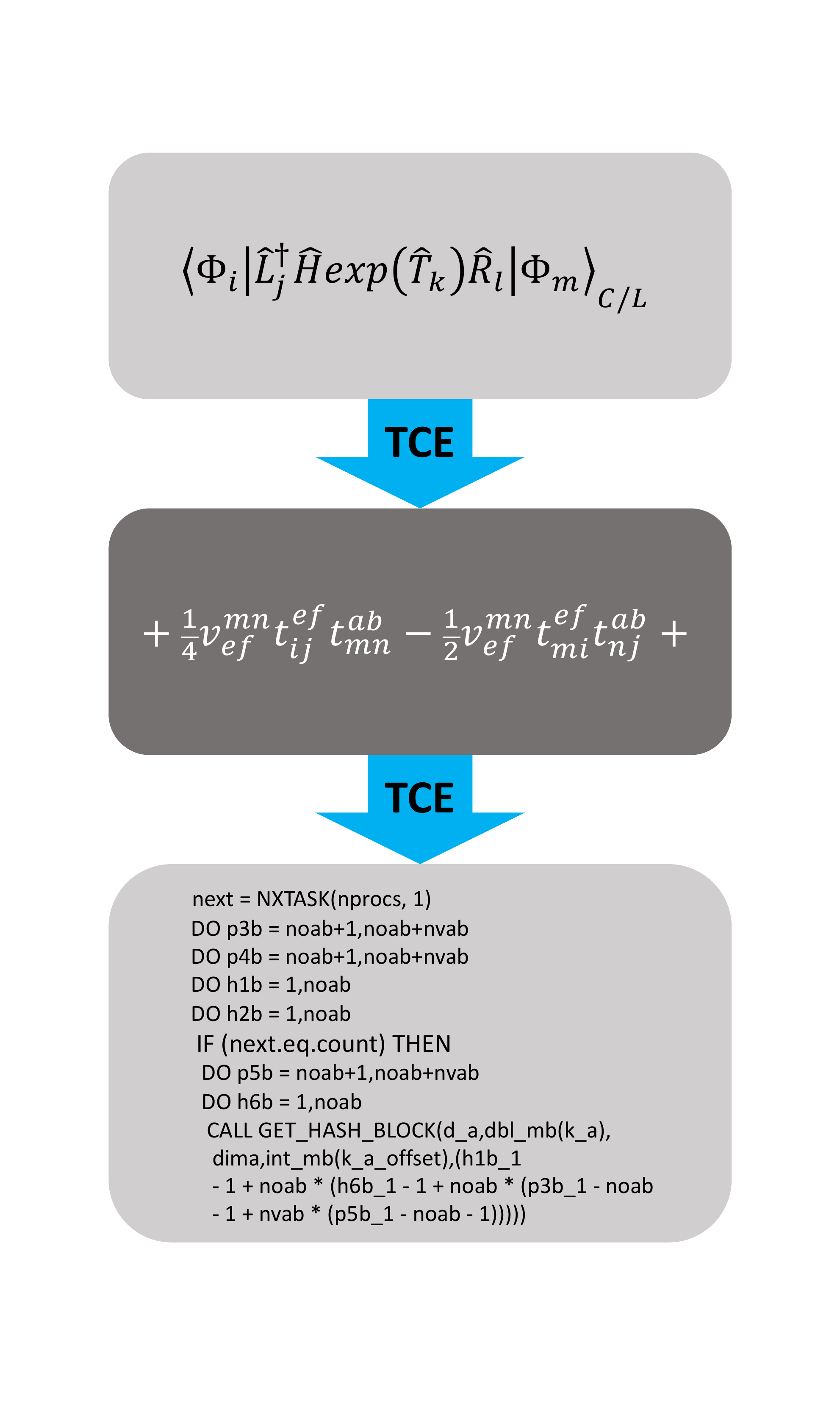}
\caption{
A schematic representation of TCE workflow (see text for details).
}
\label{tceworkflow}
\end{figure}

With the ansatz of a method given in terms of Eq.\ (\ref{TCE_eq1}), TCE first (1) evaluates these operator-determinant expressions into sums-of-products of matrices (molecular integrals and excitation amplitudes) using normal-ordered second quantization and Wick's theorem, second (2) transforms the latter into a computational sequence (algorithm), which consists in an ordered series of binary matrix multiplications and additions, and third (3) generates parallel-execution programs implementing these matrix multiplications and additions, which can be directly copied into appropriate directories of the NWChem source code and which are called by a short, high-level driver subroutine humanly written (see Fig. \ref{tceworkflow}). 

In step (2), TCE finds the (near-)minimum cost path of evaluating sums-of-products
of matrices by solving the matrix-chain problem (defining the so-called
``intermediates'') and by performing common subexpression elimination and
intermediate reuse. 
In step (3), the computer-generated codes perform dynamically-load-balanced
parallel matrix multiplications and additions, taking advantage of spin, spatial,
and index-permutation symmetries. The parallelism, symmetry usage, and memory/disk
space management are all achieved by virtue of TCE's data structure: every matrix
(molecular integrals, excitation amplitudes, intermediates, etc.)\ is split into
spin- and spatial-symmetry-adapted tiles, whose sizes are determined at runtime so
that the several largest tiles can fit in core memory. Only symmetrically-unique,
non-zero tiles are stored gapless (with their storage addresses recorded in hash
tables which are also auto-generated by TCE) and used in parallel tile-wise
multiplications and additions, which are dynamically distributed to idle processors
on a first-come, first-served basis. NWChem's parallel middleware, especially
Global Arrays, was essential for making the computer-generated parallel codes
viable. 

TCE is a part of the NWChem source-code distribution, and a user is encouraged to
implement their own ansatzes into high-quality parallel codes. 
Therefore TCE has paved the way for quick development of various implementations of coupled-cluster methods that would take 
disproportionately longer time if hand-coded. Additionally, TCE provided a new testing ground for several novel parallel algorithms 
for accurate many-body methods and has been used to generate a number of canonical implementations of single reference CC methods for 
ground- and excited-state calculations  for arbitrary reference function including: RHF, ROHF, UHF, and multi-reference cases.  
Below we listed basic components 
of the TCE infrastructure in NWChem:
\begin{itemize}
\setlength{\itemsep}{0pt}%
\setlength{\parskip}{0pt}
\item various perturbative methods ranging from second (MBPT(2)/MP2) to fourth-order (MBPT(4)/MP4) of
M\o{}ller-Plesset perturbation theory, 
\item single reference iterative (CCD,\cite{cizek66_4256} CCSD,\cite{purvis82_1910} CCSDT,\cite{ccsdt_noga,ccsdt_noga_err,scuseria_ccsdt} 
CCSDTQ \cite{ccsdtq_nevin,kucharski1991}) 
and non-iterative (CCSD(T),\cite{Raghavachari}
CR-CCSD(T),\cite{kowalski2000method} 
LR-CCSD(T),\cite{ndc1} 
CCSD(2),\cite{gwaltney2000second,gwaltney2000second,gwaltney2001second,hirata2004combined} 
CCSD(2)$_T$,\cite{hirata2004combined} 
CCSDT(2)$_Q$) \cite{hirata2004combined} 
CC approximations for ground-state calculations,
\item single reference iterative ( EOMCCSD,
\cite{geertsen1989equation,comeau1993equation}
EOMCCSDT, 
\cite{kowalski2001active,kucharski2001coupled}
EOMCCSDTQ
\cite{hirata2004higherex,kallay2004calculation}
) 
and non-iterative (CR-EOMCCSD(T)\cite{creom2004})  Equation-of-motion CC (EOMCC) approximations \cite{stanton1993equation} for 
excited-state calculations,
\item Ionization potential and electron affinity EOMCC (IP/EA-EOMCC) methods,
\cite{nooijen1995equation,stanton1995perturbative,musial2003equation,musial2003equation2,gour2006efficient,kamiya2006higher,kamiya2007higher}
\item linear-response CC (LR-CC) methods for calculating static and frequency-dependent polarizabilities and static hyperpolarizabilities at the CCSD and CCSDT levels of approximation,\cite{koch1990coupled}
\item state-specific multi-reference CC (MR-CC) methods for quasi-degenerate systems.\cite{pittnermasik,bwpittner1,mahapatra1,mahapatra2,mahapatra3,evangelista1,evangelista2} 
\end{itemize}
The TCE infrastructure has also been used in exploring new parallel algorithms and algorithms for emerging computer architectures. The most important examples include:
\begin{itemize}
\setlength{\itemsep}{0pt}%
\setlength{\parskip}{0pt}
\item parallel algorithms for excited-state CR-EOMCCSD(T) calculations with demonstrated scalability across 210,000 cores of 
Jaguar Cray XT5 system at the Oak Ridge Leadership Computing  Facility  (OLCF) \cite{kowalski2011scalable},
\item new CC algorithms   for GPU and Intel MIC architectures (single-reference CC and MR-CC theories),\cite{ma2011gpu,bhaskaran2013noniterative,apra2014efficient,apra2016implementation,bylaska2017transitioning}
\item new algorithms for multi-reference CC methods utilizing processor groups and multiple levels of parallelism 
(the so-called reference-level of parallelism of Refs.
\cite{brabec2012parallel,bhaskaran2012implementation})
with demonstrated scalability across 80,000 cores of Jaguar Cray XT5 system,\cite{bhaskaran2012implementation}
\item new execution models for the iterative CCSD and EOMCCSD models.\cite{kowalski2011scalable}
\end{itemize}
With  TCE, one can perform  CC calculations for closed- and open-shell systems characterized by 1,000-1,300 orbitals. 
Some of the most illustrative examples of  TCE calculations are (1)  static and frequency-dependent polarizabilities for the $C_{60}$ molecule,\cite{kowalski2008coupled}
excited state simulations for $\pi$-conjugated chromophores,\cite{doi:10.1021/ct200217y} and IP-EOMCCSD calculations for ferrocene with explicit inclusion of solvent molecules. 
One cutting edge application of TCE CC was the early application of  EOMCC methodologies in 
excited-state studies  of functionalized forms of porphyrin \cite{kowalski2011scalable}. 
Additionally, TCE has also served as a development platform for early implementations of the coupled-cluster Green's function formalism.\cite{nooijen92_55,nooijen93_15,nooijen95_1681,meissner93_67}
The TCE development has since been followed by several other efforts towards enabling scalable tensor libraries. This includes Super Instruction Assembly Language
SIAL,\cite{deumens2011super,deumens2011software}  Cyclop Tensor Framework (CTF), \cite{solomonik2014massively} TiledArray framework,\cite{peng2016massively} and Libtensor,\cite{epifanovsky2013new} 
which have been used to develop scalable implementations of CC methods.

\subsection{Relativistic Methods} 
\label{relativistic}
Methods which include treatment of relativistic effects are
based on the Dirac equation\cite{dirac1928}, which has a four-component
wavefunction. The solutions to the Dirac equation describe both positrons (the ``negative energy'' states) and electrons (the ``positive energy'' states), as well as both spin orientations, hence the four components. The wavefunction may be broken down into two-component functions traditionally known as the large and small components; these
may further be broken down into the spin components.\cite{dyall2007book,saue2011primer,Reiher15_book, autschbach2015book}

The implementation of approximate all-electron relativistic methods in
quantum chemical codes requires the removal of the negative energy
states and the factoring out of the spin-free terms. Both of these may
be achieved using a transformation of the Dirac Hamiltonian known in
general as a Foldy-Wouthuysen (FW) transformation. Unfortunately, this
transformation cannot be represented in closed form for a general
potential, and must be approximated. One popular approach is the
Douglas and Kroll\cite{rel_1} method developed by
Hess\cite{rel_2,rel_3}. This approach decouples the positive and
negative energy parts to second-order in the external potential (and
also fourth-order in the fine structure constant, $\alpha$). Other
approaches include the zeroth order regular approximation
(ZORA)\cite{rel_4,rel_5,rel_6,nichols2009}, modification of the Dirac
equation by Dyall\cite{rel_8}, which involves an exact FW
transformation on the atomic basis set level\cite{rel_9,rel_10} and
the exact 2-component (X2C) formulation, which is a catch-all for a
variety of methods that arrive at an exactly decoupled two-component
Hamiltonian using matrix
algebra.\cite{peng2012x2c,autschbach2012x2c,autschbach2014relativistic,autschbach2015book,autschbach2017x2c}
NWChem contains released implementations of the DKH, ZORA, and Dyall
approaches, while the X2C method is available in a development
version.\cite{autschbach2012x2c,autschbach2017x2c}

Since these approximations only modify the integrals, they can, in principle, be used at all levels of theory. At present, the Douglas-Kroll, ZORA and X2C implementations can be used at all levels of theory, whereas Dyall's approach is currently available at the Hartree-Fock level. 

\paragraph{Douglas-Kroll Approximation:}

NWChem contains three second-order Douglas-Kroll approximations termed as FPP, DKH, and
DKHFULL. The FPP is the approximation based on free-particle
projection operators\cite{rel_2}, whereas the DKH and DKFULL
approximations are based on external-field projection
operators\cite{rel_3}. The latter two are considerably better
approximations than the former. DKH is the Douglas-Kroll-Hess approach
and is the approach that is generally implemented in quantum chemistry
codes. DKFULL includes certain cross-product integral terms ignored in
the DKH approach (see for example, H{\"a}berlen and
R{\"o}sch\cite{rel_13}). The third-order Douglas-Kroll approximation (DK3) 
implements the method by Nakajima and Hirao\cite{rel_14,rel_15}.

\paragraph{Zeroth Order Regular Approximation (ZORA):}
The spin-free and spin-orbit  versions of the one-electron zeroth order regular approximation (ZORA) have been implemented. Since the ZORA correction depends on the potential, it is not gauge invariant. This is addressed by using the atomic approximation of van Lenthe and coworkers.\cite{van2000gradients,van2006note} Within this approximation, the ZORA corrections are calculated using the superposition of densities of the atoms in the system. As a result, only intra-atomic contributions are involved, and no
gradient or second derivatives of these corrections need to be calculated. 
In addition, the corrections need only to be calculated once at the start of the calculation and stored. The ZORA approach is implemented in two ways in NWChem, one where the ZORA potential components are directly computed on an all-electron grid\cite{nichols2009} and a second approach, where the ZORA potential is computed using the model potential approach due to van W{\"u}llen and co-workers.\cite{rel_16,rel_17} 

\paragraph{Dyall's Modified Dirac Hamiltonian Approximation:}
The approximate methods described in this section are all based on Dyall's modified Dirac Hamiltonian. This Hamiltonian is entirely equivalent to the original Dirac Hamiltonian, and its solutions have the same properties. The modification is achieved by a transformation on the small component. This gives the modified small component the same symmetry as the large component. The advantage of the modification is that the operators now resemble those of the Breit-Pauli Hamiltonian, and can be classified in a similar fashion into spin-free, spin-orbit, and spin-spin terms. It is the spin-free terms which have been implemented in NWChem, with a number of further approximations. Negative energy states are removed by a normalized elimination of the small component (NESC), which is equivalent to an exact Foldy-Wouthuysen (EFW) transformation. Both one-electron and two-electron versions of NESC (NESC1E and NESC2E, respectively) are available, and both have analytic gradients.\cite{rel_8,rel_9,rel_10}

\subsection{Molecular Properties}
A broad array of simple and response-based molecular properties can be
calculated using the HF and DFT wavefunctions in NWChem. These
include: natural bond analysis, dipole, quadrupole, octupole moments,
Mulliken population analysis and bond order analysis, L{\"o}wdin
population analysis, electronic couplings for electron transfer,\cite{faradzel1990,rosso2003}, Raman spectroscopy,\cite{schatz-1,schatz-2},
electrostatic potential (diamagnetic shielding)
at nuclei, electric field and field gradient at nuclei, electric field
gradients with relativistic effects\cite{autschbach-7}, electron and spin density at
nuclei, GIAO-based NMR properties like shielding, hyperfine coupling
(Fermi-Contact and Spin-Dipole expectation values), indirect
spin-spin coupling,\cite{ditchfield-giao,dupuis-giao,autschbach-1} G-shift,\cite{autschbach-3} EPR, paramagnetic NMR parameters,\cite{autschbach-0,autschbach-2} and optical activity.\cite{autschbach-4,autschbach-5,autschbach-6,srebro2011}
Note that only linear-response is supported and for single frequency, electric field, and mixed electric-magnetic field perturbations. Ground state and dynamic dipole polarizabilities for molecules can be calculated at the CCSD, CCSDT, and CCSDTQ levels using the linear-response formalism.\cite{hammond2009accurate} For additional information, we refer the reader to the online manual.\cite{nwchem.org}

\subsection{Periodic Plane-Wave Density Functional Theory}
\label{pwdft}
The NWChem plane-wave density functional theory (NWPW) module contains two programs:
\begin{itemize}
    \item PSPW - A pseudopotential and projector augmented (PAW) plane-wave $\Gamma$-point code for calculating molecules, liquids, crystals, and surfaces,
    \item BAND - A pseudopotential plane-wave band structure code for calculating crystals and surfaces with small band gaps (e.g. semi-conductors and metals),
\end{itemize}
These programs use a common infrastructure for carrying out
operations related to plane-wave basis sets that is parallelized with
the MPI and OpenMP libraries\cite{Mathias17,bylaskaintel17,bylaska2017transitioning,bylaska2002parallel,bylaska2011large,bylaska2011parallel,bylaska2017plane,ValievBylaskaWeare2002}
The NWPW module can
be used to carry out many different kinds of simulations.  In addition
to the standard simulations implemented in other modules, e.g. energy,
optimize, and freq, there are additional capabilities specific to PSPW
and BAND that can be used to carry out NVE and NVT~\cite{blochl1992adiabaticity} Car–Parrinello~\cite{car1985unified} and
Born-Oppenheimer molecular dynamics simulations, hybrid {\it ab initio}
molecular dynamics and molecular mechanics (AIMD-MM)
simulations~\cite{cauet_2012,bylaska2017plane}, Gaussian/Fermi/Marzari-Vanderbilt smearing,
Potential-of-Mean-Force (PMF)~\cite{roux1995calculation}/Metadynamics~\cite{bussi2006equilibrium,barducci2011metadynamics}/Temperature-Accelerated-Molecular-Dynamics (TAMD)~\cite{maragliano2008single,maragliano2006temperature}/Weighted-Histogram-Analysis-Method (WHAM)~\cite{kumar1992weighted} free energy simulations, AIMD-EXAFS
simulations using open-source versions of the FEFF software~\cite{rehr2001progress,rehr1990scattering,ankudinov1998real} that have
been parallelized, electron transfer calculations~\cite{bylaska2018corresponding}, unit cell
optimization, optimizations with space group symmetry, Monte-Carlo NVT
and NPT simulations, phonon calculations, simulations with spin-orbit
corrections, Wannier~\cite{silvestrelli1999maximally} and rank reducing density matrix~\cite{damle2015compressed} localization
calculations, Mulliken~\cite{kawai1991instability} and Bl{\"o}chl~\cite{blochl1995electrostatic} charge analysis, Gaussian cube file
generation, periodic dipole and infrared (AIMD-IR) simulations, band
structure plots, density of states.  Calculations can also be run
using a newly developed i-PI~\cite{kapil2019ipi} interface, and more direct interfaces to
ASE~\cite{Hjorth_Larsen_2017}, nanoHUB~\cite{klimeck2008nanohub}, and EMSL Arrows~\cite{emslarrows} simulation tools are currently being implemented.

A variety of exchange-correlation functionals have been implemented in
both codes, including the local density approximation (LDA)
functionals, generalized gradient approximation (GGA) functionals,
full Hartree-Fock and screened exchange, hybrid DFT functionals,
self-interaction correction (SIC) functionals~\cite{bylaska2006new}, localized exchange
method, DFT+U method, and Grimme dispersion corrections~\cite{grimme2010,grimme2011}, as well as
recently implemented vdW dispersion functionals~\cite{langreth2005van}, and meta-generalized
gradient approximation (metaGGA) functionals. The program contains
several codes for generating pseudopotentials, including Hamann~\cite{hamann1989generalized} and
Troulier-Martin~\cite{troullier1991efficient}, and PAW~\cite{blochl1994projector} potentials. These codes have the option for
generating potentials with multiple projectors and semi-core
corrections.  It also contains codes for reading in HGH~\cite{hartwigsen1998relativistic}, GTH~\cite{goedecker1996separable}, and
norm-conserving pseudopotentials in the CPI and TETER formats. Codes
for reading Optimized Norm-Conserving Vanderbilt (ONCV) pseudopotentials~\cite{hamann2013optimized,schlipf2015optimization} and USPP PAW potentials will become
available in future releases of NWChem.

\begin{figure}[ht]
\centering
\includegraphics[scale=0.25]{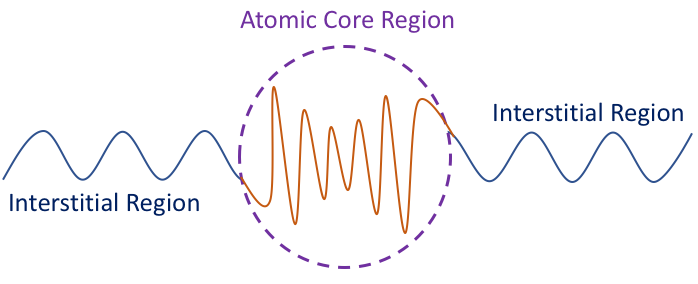}
\caption{Illustration of the atomic core and interstitial regions in a valence wavefunction.  Bonding takes place in the interstitial region and the atomic core regions change very little from molecule to molecule. Figure from Bylaska {\it et al.}~\cite{bylaska2017plane}.}
\label{fig.interstitial}
\end{figure}

The pseudopotential plane-wave DFT methods implemented in NWChem are a fast and efficient way to calculate molecular and solid-state properties using DFT~\cite{hohenberg1964inhomogeneous,kohn1965self,parr1994density,pickett1989electronic,ihm1979momentum,car1985unified,payne1992iterative,remler1990molecular,kresse1996efficient,marx2000modern,martin2004electronic,ValievBylaskaWeare2002,bylaska2011large,chen2016first}. In these approaches, the fast varying parts of the valence wavefunctions inside the atomic core regions and the atomic core wavefunctions are removed and replaced by pseudopotentials~\cite{phillips1958energy, phillips1959new, austin1962general, yin1982theory,  bachelet1982pseudopotentials, hamann1989generalized, troullier1991efficient}. Pseudopotentials are chosen such that the resulting pseudoatoms have the same scattering properties as the original atoms. The rationale for this approach is that the changes in the electronic structure associated with making and breaking bonds only occur in the interstitial region outside the atomic core regions (see Fig.~\ref{fig.interstitial}).  Therefore, removing the core regions should not affect the bonding of the system. For this approach to be useful, it is necessary for the pseudopotentials to be smooth in order for plane-wave basis sets to be used.  As the atomic potential becomes stronger the core region becomes smaller and the pseudopotential grows steep. As a result, the pseudopotential can become very stiff, requiring large plane-wave basis sets (aka cutoff energies), for the first-row transition metals atoms, the lanthanide atoms, and towards the right-hand side of the periodic table (fluorine).  

The projected augmented plane-wave method (PAW)~\cite{blochl1994projector, holzwarth1997comparison, kresse1999ultrasoft, valiev1999projector, bylaska2002parallel} is another related approach that removes many of the problems of the somewhat ad hoc nature of the pseudopotentials approach. However, in the PAW approach, instead of discarding the rapidly varying parts of the electronic functions, these are projected onto a local basis set (e.g., a basis of atomic functions), and no part of the electron density is removed from the problem.  Another key feature of PAW is that by maintaining a local description of the system, the norm-conservation condition (needed for proper scattering from the core) can be relaxed, which facilitates  the use of smaller plane-wave basis sets (aka cutoff energies) then for many standard pseudopotentials.  Historically, the PAW method was implemented as a separate program in the NWPW module, rather than being fully integrated into the PSPW and BAND codes.  This separation significantly hindered its development and use.  As of NWChem version 6.8 (released in 2017), the PAW approach has been integrated into the PSPW code, and it is currently being integrated into the BAND code.  It will become available in future releases of NWChem.

\begin{figure}[ht]
\centering
\includegraphics[scale=0.55]{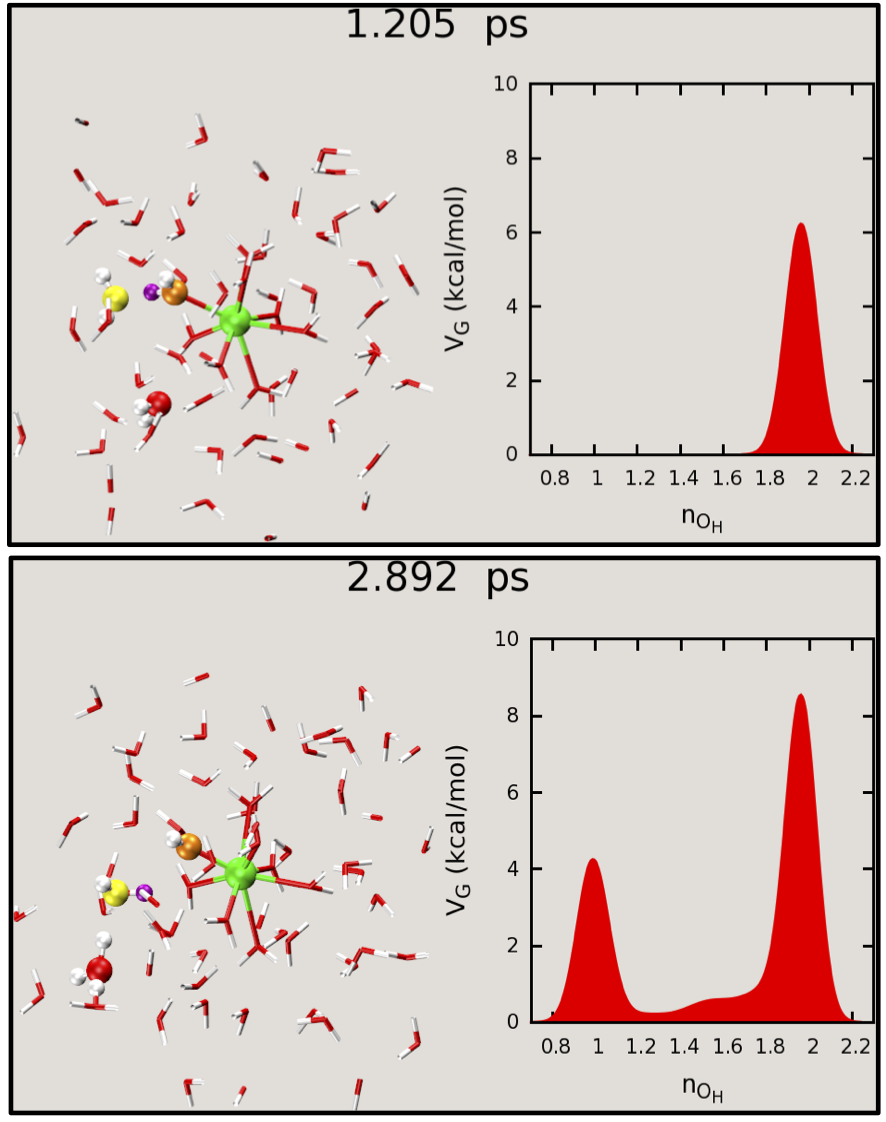}
\caption{Snapshots from a metadynamics simulation of the  hydrolysis of the U$^{4+}$ aqua ion~\cite{atta2012free}. During the simulation a proton jumps from a first shell water molecule to a second shell water molecule and then subsequently to other water molecules via a Grotthuss mechanism.}
\label{fig.freeenergy}
\end{figure}

In recent years, with advances in High-Performance Computing (HPC) algorithms and computers, it is
now possible to run AIMD simulations up to $\sim$1 ns for non-trivial
system sizes.  As a result, it is now possible to effectively use
free-energy methods with AIMD and AIMD/MM approaches.  Free energy
approaches are useful for simulating reactions where traditional
quantum chemistry approaches can be difficult to use and often require
the expertise of a very experienced quantum chemist, e.g. reactions
that are complex with concerted or multi-step components and/or
interact strongly with the solvent. Recent examples include solvent
coordination and hydrolysis of actinides metals\cite{solve2,solve1,autschbach2015book,thorium_2016,atta2012free} (see Fig.~\ref{fig.freeenergy}), hydrolysis of
explosives~\cite{bylaska2017plane}, and ion association in AlCl$_3$\cite{cauet_2012}.  To help users learn how
to use these new techniques, we developed a tutorial on carrying out
finite temperature free energy calculations in NWChem\cite{wham_tutorial}.

The NWPW module continues to be actively developed.  There are
on-going developments for RPA and GW-RPA methods, an electron transfer
MCSCF method, Raman and M\"{o}ssbauer spectroscopy, and a hybrid
method that integrates classical DFT~\cite{meng2014numerical} into {\it ab initio} molecular dynamics
(AIMD-CDFT).  In addition to these developments, we are actively
developing the next generation of plane-wave codes as part of the NWChemEx project.
These
new codes, which are being completely written from scratch, will
contain all the features currently existing in the NWPW
module.  Besides implementing fast algorithms to use an even larger
number of cores and new algorithms to run efficiently on GPUs, it
includes a more robust infrastructure to facilitate the implementation
of an $O(N)$ DFT code based on the work of Fattebert {\it et
al}.\cite{Fattebert_2014}

\subsection{Optimization, Transition State, and Rate theory Approaches}
A variety of drivers and interfaces are available in NWChem to perform
geometry minimization and transition state optimizations.  The default
algorithms in NWChem for performing these optimizations are
quasi-Newton methods with line searches.  These methods are fairly
robust, and they can be used to optimize molecules, clusters, and
periodic unit cells and surfaces.  They can also be used in
conjunction with both point group and space group symmetries, excited
state TDDFT surfaces, as well as with a variety of external fields,
such as external point charges,
COSMO\cite{klamt1993} or
SMD\cite{marenich2009}  Model.
The default methods also work
seamlessly with electronic structure methods that do not have nuclear
gradients implemented by automatically using finite difference
gradients.  NWChem also contains default methods for calculating
harmonic vibrational frequencies and phonon spectra for periodic
systems. These methods are able to make use of analytic Hessians if
they are available, otherwise a finite difference approach is used.  A
vibrational self-consistent field\cite{vscf_1999} (VSCF) method is
also available in NWChem and it can be used to calculate anharmonic
contributions to specified vibrational modes.  There is also an interface called DIRDYVTST~\cite{DIRDYVTST} that uses NWChem to compute energies, gradients, and Hessians for direct dynamics calculations with POLYRATE~\cite{POLYRATE}.

A variety of external packages, such as ASE\cite{Hjorth_Larsen_2017,ASE} and
Sella\cite{sella_2019,sella}, can also be used for
finding energy minima, saddle points on energy surfaces, and frequencies using 
either python scripting or a newly developed i-PI\cite{kapil2019ipi} interface. 
Python programs may be directly embedded into
the NWChem input and used to control the execution of NWChem. The
python scripting language provides useful features, such as variables,
conditional branches, and loops, and is also readily extended. Other
example applications for which it could be used include scanning
potential energy surfaces, computing properties in a variety of basis
sets, optimizing the energy with respect to parameters in the basis set,
computing polarizabilities with a finite field, simple molecular
dynamics, and parallel in time molecular dynamics~\cite{bylaska2013extending}.

NWChem also contains an implementation of the nudged elastic band
(NEB) method of J{\'o}nsson and co-workers\cite{neb_1998,neb_2000a,henkelman2000climbing,neb_2014}
and the zero-temperature string method
of vanden Eijden {\it et al}.\cite{doi:10.1063/1.2720838}
Both these methods can be used to 
find minimum energy paths.  Currently, a quasi-Newton
algorithm is used for the NEB optimization. A better approach for this
kind of optimization is to use a non-linear multi-grid algorithm, such
as the Full Approximation Scheme (FAS)~\cite{henson2002multigrid}.  A new implementation of NEB
based on FAS is available on Bitbucket\cite{python_neb}, and an
integrated version will soon be available  in NWChem.

\subsection{Classical Molecular Dynamics}
\label{cmd}
\begin{figure*}[t]
\centering
\includegraphics[scale=0.38]{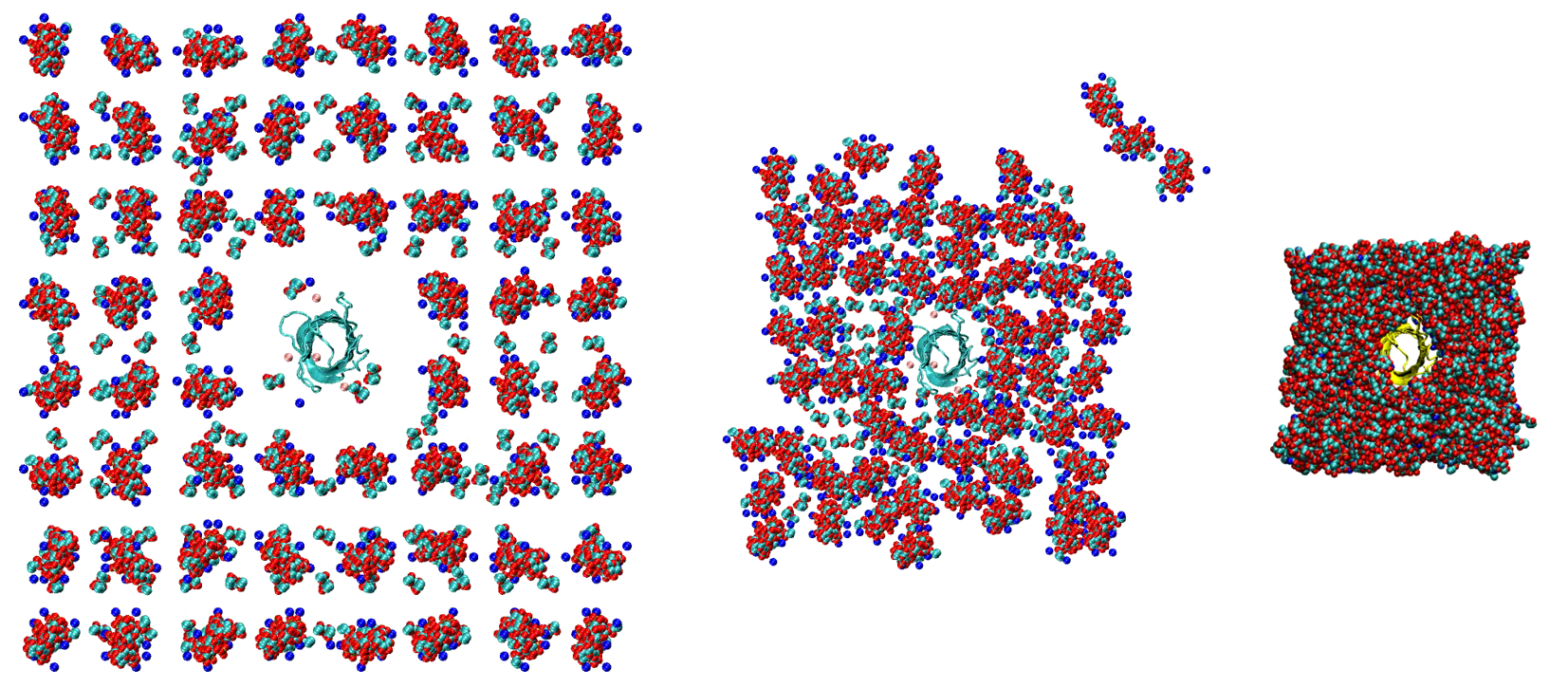}
\caption{
The NWChem MD Prepare utility facilitates the setup of trans-membrane proteins in complex asymmetric membrane environments in a semi-automated procedure. Shown here are the top views of step 1 in which membrane lipopolysaccharide molecules with the necessary counter ions are placed on a rectangular grid around a trans-membrane protein, in which each membrane lipid molecule is randomly rotated around the principal molecular axis (left panel), step 2 in which each cluster of a lipid molecule is translated towards the center of the transmembrane protein such that no steric clashed occur (center panel), and step 3 in which the system is equilibrated using strict restraint potentials to keep the lipid molecules aligned along the normal of the membrane and the lipid head groups in the plane of the membrane (right panel). After this procedure, the system would be solvated and equilibrated while slowly removing the positional restraint potentials.
}
\label{membrane}
\end{figure*}

The integration of a molecular dynamics (MD) module in NWChem enables the generation of time evolution trajectories based on Newton’s equation of motion of molecular systems in which the required forces can originate from a classical force field, any implemented quantum mechanical method for which spatial derivatives have been implemented, or hybrid quantum mechanical/molecular mechanical (QM/MM) approaches. The method is based on the ARGOS molecular dynamics software, originally designed for vector processors,\cite{Straatsma1990} but later redesigned for massively parallel architectures.\cite{Straatsma2000a,Straatsma2005a,Straatsma2011,Straatsma2013}
\\
\paragraph{System Preparation:}
The preparation of a molecular system is done by a separate prepare module that reads the molecular structure and assembles a topology from the databases with parameters for the selected force field. The topology file contains all static information for the system. In addition, this module generates a so-called restart file with all dynamic information. The prepare module has a wide range of capabilities that include the usual functions of placing counter-ions and solvation with any solvent defined in the database. The prepare module is also used to define Hamiltonian changes for free energy difference calculations, and the definition of those parts of the molecular systems that will be treated quantum mechanically in QM/MM simulations. Some of the more unique features include setting up a system for proton hopping (QHOP) simulations,\cite{Straatsma2000,Gu2007} and the setup of biological membranes from a single lipid-like molecule. This last capability has been successfully used for the first extensive simulation studies of complex asymmetric lipopolysaccharide membranes of Gram-negative microbes\cite{ Lins2001,Shroll2002,Straatsma2005,Soares2008a,Soares2008b} and their role in the capture of recalcitrant environmental heavy metal ions,\cite{Lins2008} microbial adhesion to geochemical surfaces,\cite{Shroll2003a,Shroll2003b,Straatsma2003,Felmy2005} and the structure and dynamics of trans-membrane proteins including ion transporters \cite{Straatsma2006,Soares2007,Straatsma2009}(Fig. ~\ref{membrane}).
\\

\paragraph{Force Fields:}
The force field implemented in NWChem consists of harmonic terms for bonded, angle and out of plane bending interactions, and trigonometric terms for torsions. Non-bonded van der Waals and electrostatic interactions are represented by Lennard-Jones and Coulombic terms, respectively. Non-bonded terms are evaluated using charge groups and subject to a user-specified cutoff radius. Electrostatic interaction corrections beyond the cutoff radius are estimated using the smooth particle mesh Ewald method.\cite{Essmann1995} Parameter databases are provided for the AMBER\cite{Ponder2003} and CHARMM\cite{Vanommeslaeghe2012} force fields.

Even for purely classical MD simulations, the integration with the electronic structure methods provides a convenient way of determining electrostatic parameters for missing fragments in standard force field databases, through the use of restrained electrostatic potential fitting\cite{Bayly1993,Cornell1993} to which a variety of additional constraints and restraints can be applied.\\

\paragraph{Simulation Capabilities:}

Ensemble types available in NWChem are NVE, NVT, and NPT, using the Berendsen thermostat and barostat.\cite{Berendsen1984} Newton's equations of motion are integrated using the standard leap-frog Verlet or velocity Verlet algorithms. A variety of fundamental properties are evaluated by default during any molecular dynamics simulation. Parallel execution time analysis is available to determine the parallel efficiency.

The MD module has extensive free energy simulation capabilities,\cite{Straatsma1991a,Straatsma1991b,Straatsma1992,Straatsma1993,Straatsma1995,Straatsma1999}
which are implemented in a so-called multi-configuration approach. For each incremental change of the Hamiltonian to move from the initial to the final state, sometimes referred to as a window, a full molecular simulation is carried out. This allows for a straightforward evaluation of statistical and systematic errors where needed, including a correlation analysis.\cite{Straatsma1986}
Based on the ARGOS code\cite{Straatsma1990} it has some unique features, such as the separation-shifted scaling technique to allow atoms to appear from or disappear to dummy atoms.\cite{Zacharias1994} 
One of the advantages of the integration of MD into the electronic structure methods framework in NWChem is the ability to carry out hybrid QM/MM simulations (discussed in the next section). The preparation of molecular systems for the MD module allows for flexibly specifying parts of the molecular system to be treated by any of the implemented electronic structure methods capable of evaluating positional gradients.

A unique feature in the NWChem MD module is the optional specification of protonatable sites on both solute and solvent molecules. Pairs of such sites can dynamically change between protonated or unprotonated state, effectively exchanging a proton. Transitions are governed by a Monte Carlo type stochastic method to determine when transitions occur. This so-called QHOP approach was developed by the research group of Helms.\cite{Gu2007}\\

\paragraph{Analysis Capabilities:}
The NWChem MD capability includes two analysis modules. The original analysis module, {\it analyze}, analyzes trajectories in a way that reads individual structures one time step at a time and distributes the data in a domain decomposition fashion as in the molecular simulation that generated the data. The second data-intensive analysis module, {\it diana}, reads entire trajectories and distributes the data in the time domain. This is especially effective for analyses that require multiple passes through a trajectory, but requires the availability of potentially large amounts of memory. 
\cite{Straatsma2007,Peterson2008} An example of such analyses is the Essential Dynamics Analysis, a principal component analysis (PCA) based calculation to determine the dominant motions in molecular trajectories.\\

\paragraph{Parallel Implementation Strategy:}
The most effective way of distributing a system with large numbers of
particles is through the use of domain decomposition of the physical
space. The implementation in NWChem, facilitated through the use of
the Global Arrays (GA) toolkit, partitions the simulation space into
rectangular cells that are assigned to different processes ranks or
threads. Each of these ranks carries out the calculation of intra-cell
atomic energies and forces of the cells assigned. Inter-cell energies
and forces are evaluated by one of the ranks that was assigned one or
the other of the cell pairs.

Two load balancing methods have been implemented in NWChem, both based
on measured computation time. In the first one, the assignment of
inter-node cell pair calculations is redefined such that assignments
move from the busiest node to the less busy node. This scheme requires
minimal additional communication, and since only two nodes are involved
in the redistribution of work, the communication is local, i.e. node
to node. In the second scheme, the physical size of the most
time-consuming cell is reduced, while all other cells are made
slightly larger. This scheme requires communication and redistribution
of atoms on all nodes. In practice, the first scheme is used until
performance no longer improves, after which the second scheme is used
once followed by returning to use the first scheme. This approach has
been found to improve load balancing even in systems with a very
asymmetric distribution of computational
intensity.\cite{Straatsma2001}



\section{Hybrid methods}
We define hybrid methods as those coupling different levels of
description to provide an efficient calculation of a chemical system,
which otherwise may be outside the scope of conventional single-theory
approaches. The physical motivation for such methods rests on the
observation that, in the majority of complex chemical systems, the
chemical transformation occurs in localized regions surrounded by an
environment, which can be considered chemically inert to a reasonable
approximation.
Since hybrid methods require the combination of multiple theoretical methods in a single
simulation, the diversity of simulation methodologies available in NWChem makes it a platform particularly apt for this purpose.

One common example involves chemical transformations in a bulk solution environment, forming the foundations of wide variety of spectroscopic measurements (UV-vis, NMR, EPR, etc.). The reactive region, referred to as the ``solute'', involves electronic structure degrees of freedom and thus requires the quantum mechanical (QM) based description, such as DFT or more complex wavefunction methods. In the conventional approach, such QM description would be necessarily extended to the entire system making the problem a heroic computational task. In a hybrid approach, the treatment of a surrounding environment ("solvent") would be delegated to a much simpler description, such as the continuum model (CM), for example.
The latter is supported in NWChem via two models - COSMO\cite{klamt1993}
(COnductor-like Screening MOdel) and SMD\cite{marenich2009} (Solvation Model based on
Density) Model. The resulting QM/CM approaches are particularly well
suited for accurate and efficient calculation of solvation free
energies, geometries in solution, and spectroscopy in solution.  The
SMD model employs the Poisson equation with non-homogeneous dielectric
constant for bulk electrostatic effects, and
solvent-accessible-surface tensions for cavitation, dispersion, and
solvent-structure effects, including hydrogen bonding. For
spectroscopy in solution, the Vertical Excitation (or Emission) Model
(VEM) has also been implemented for calculating the vertical
excitation (absorption) or vertical emission (fluorescence) energy in
solution according to a two-time-scale model of solvent
polarization\cite{marenich2011-vem}.

For systems where an explicit solvation environment treatment is
needed (for example, heterogeneous systems like a protein matrix),
NWChem provides a solution in terms of combined quantum mechanics/molecular mechanics
(QM/MM)  approach.\cite{Valiev2007,Valiev2008} Here, the environment is
described at the classical molecular mechanics level. This offers more
fidelity compared with a continuum solvent description, while still keeping the computational costs down. The total energy of the system in QM/MM approach can be represented as a sum of the energies corresponding to QM and MM regions:
\begin{equation}
    E(\mathbf{r}, \mathbf{R} ; \psi)=E_{q m}(\mathbf{r}, \mathbf{R} ; \psi)+E_{m m}(\mathbf{r}, \mathbf{R})
\end{equation}
where $\psi$ denotes electronic degrees of freedom, and $\mathbf{r}, \mathbf{R}$ refer to nuclear coordinates of QM and MM regions correspondingly.
The QM energy term can be further decomposed into internal and external parts
\begin{equation}
    E_{q m}[\mathbf{r}, \mathbf{R} ; \psi]=E_{q m}^{\mathrm{int}}[\mathbf{r}; \psi]+E_{q m}^{e x t}[\mathbf{r}, \mathbf{R} ; \rho]
\end{equation}
where $\rho$ is the electron density.

As a generic module, the QM/MM implementation can utilize any of the Gaussian basis set based QM
modules available in NWChem and supports nearly all the task
functionalities. The calculation of QM energy remains the main
computational expense in the QM/MM approach. This issue is more pronounced compared with the continuum coupling case, because of the additional atomistic degrees of freedom associated with MM description. The latter comes into play
because any change in the MM degrees of freedom will, in general,
trigger the recalculation of the QM energy
($E_{qm}(\mathbf{r}, \mathbf{R} ; \psi)$). To alleviate these issues during
the optimization, the QM/MM module offers the option of alternating
relaxation of QM and MM regions. During the latter phase, the user may
utilize an approximation where the QM degrees of freedom are kept frozen
until the next cycle of QM region relaxation, offering significant
computational savings. A similar technique can be utilized in the
dynamical equilibration of the MM region and calculations of reaction
pathways and free energies.
In addition to the native MD module, the NWChem QM/MM module can also utilize the external AMBER MD code\cite{amber2019} for running the classical part of the calculations. In this case, QM/MM simulations involve two separate NWChem and AMBER calculations with data exchange mediated through files written to disk. 

Additionally, the QM/MM capability in NWChem has resulted in the development and refinement of force-field parameters, that can, in turn, be used in classical molecular dynamics simulations. Over the last two decades, classical parameters obtained using NWChem have been employed to
address the underlying mechanisms of a variety of novel complex
biological systems and their interactions (e.g., lipopolysaccharide
membranes, carbohydrate moieties, mineral surfaces, radionuclides,
organophosphorous compounds)\cite{Lins2001,Lins2008,Soares2008b,
doi:10.1021/ct700024h,Shroll2002,Shroll2003a,doi:10.1021/jp208787g,
doi:10.1021/jp9083635,doi:10.1021/es702688c,doi:10.1021/jp906124a}
which has led to a significant expansion
of the database of AMBER- and Glycam-compatible force fields and the GROMOS force field for lipids,
carbohydrates and nucleic acids.\cite{Chandrasekhar2003,doi:10.1002/jcc.20275,doi:10.1021/ct300479h,
doi:10.1002/jcc.23721,doi:10.1002/jcc.20193,doi:10.1002/bip.21602,
doi:10.1021/ct8002964}

For cases where a classical description of the environment is deemed insufficient, NWChem offers an option to perform an ONIOM type calculation.\cite{morokuma1996} The latter differs from QM/MM in that the lower level of theory is not restricted to its region but also encompasses regions from all the higher levels of description. For example, in the case of the two-level description, the energy is written as
\begin{equation}
    E(\mathbf{R}) =E^L(\mathbf{R})+(E^H(\mathbf{R}^H)-E^L(\mathbf{R}^H))
\end{equation}
where subscripts $H, L$ refer to high and low levels of theory correspondingly. The high-level treatment is restricted to a smaller portion of the system ($\mathbf{R}^H$), while the low level of theory goes over the entire space ($\mathbf{R}$). The second term in the above equation takes care of overcounting. The NWChem ONIOM module implements two- and three-layer ONIOM models for use in energy, gradient, geometry optimization, and vibrational frequency calculations with any of the pure QM methods within NWChem.

A new development in hybrid method capabilities of NWChem involves classical density functional theory (cDFT).\cite{valiev2020,valiev2018,valiev2012} The latter represents a classical variant of electronic structure DFT, where the main variable is the classical density of the atoms.\cite{chandler1986A,chandler1986B} Conceptually, this type of description lies between continuum and classical force field models, providing orders of magnitude improvements over classical MD simulations. The approach is based on incorporating important structural features of the environment in the form of classical correlation functions. This allows for efficient and reliable calculations of thermodynamical quantities, providing an essential link between electronic structure description at the atomistic level and phenomena observed at the macroscopic scale.

\section{Parallel Performance}
The design and development of NWChem from the outset was driven by
parallel scalability and performance to enable large scale
calculations and achieve fast time-to-solution by using many CPUs
where possible. The parallel tools outlined in
section \ref{section:parallel} provided the programming framework for
this.

The advent of new architectures such as the
GPU\cite{gpu_owensluebke2007} platforms have required the parallel coding
strategy within NWChem to be revisited. At present, the coupled-cluster
code within TCE
can utilize both the CPU
and GPU hardware at a massive scale\cite{ma2011gpu,gawande_2019}.
The emergence of many-core processors in the last ten years provided the opportunity for
starting a collaborative effort with Intel corporation to optimize NWChem on
this new class of computer architecture. As part of this collaboration,
the TCE implementation of the CCSD(T) code was ported to the Intel Xeon
Phi line of many-core processors \cite{apra2014efficient} using a
parallelization strategy based on a hybrid GA-OpenMP approach. 
The \textit{ab initio} plane-wave molecular dynamics code
(section \ref{pwdft}) has also been optimized to take full advantage of
these Intel many-core processors\cite{Mathias17,bylaskaintel17}.

In the rest of this section, we will discuss the parallel scalability and performance of the main capabilities in NWChem.

\paragraph{Gaussian Basis Density Functional Theory:}

\begin{figure}[ht] 
  \includegraphics[trim=0.33in 0.38in 0.33in 0.38in ,scale=0.35]{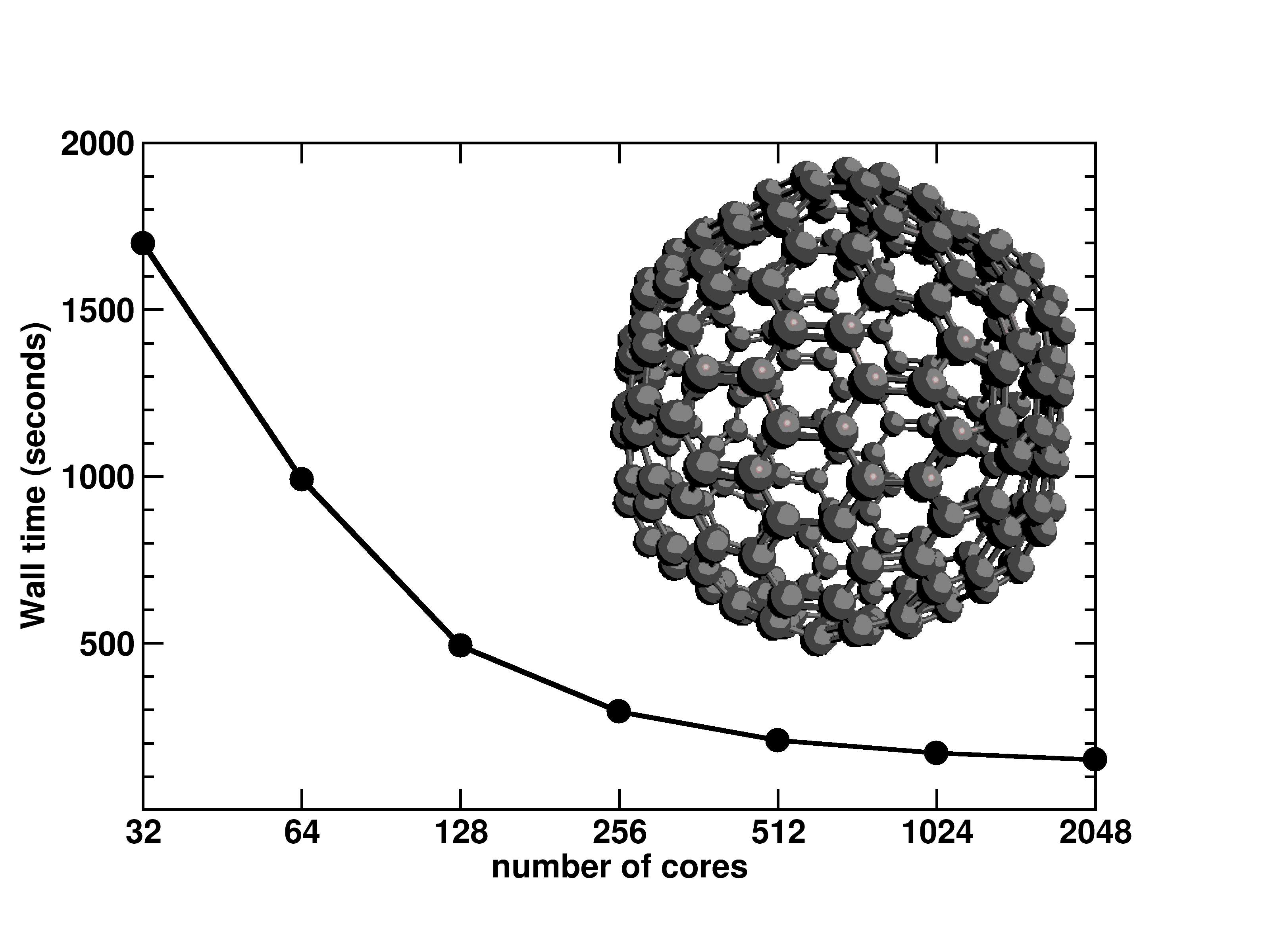}
\caption{C$_{240}$ DFT benchmark.}
\label{fig:dftc240}
\end{figure}

In Fig.~\ref{fig:dftc240}, we report the parallel performance of the Gaussian basis set DFT module in NWChem.
This calculation involved performing a PBE0 energy calculation (four SCF iterations in direct mode) on
the C$_{240}$ molecule with the 6-31G* basis set (3600 basis functions)
without symmetry. These calculations were performed on the Cascade
supercomputer located at PNNL.\\

\iftrue
\paragraph{Time-Dependent  Density Functional Theory:}

\begin{figure}[ht] 
\includegraphics[trim=0.35in 0.38in 0.33in 0.38in ,scale=0.34]{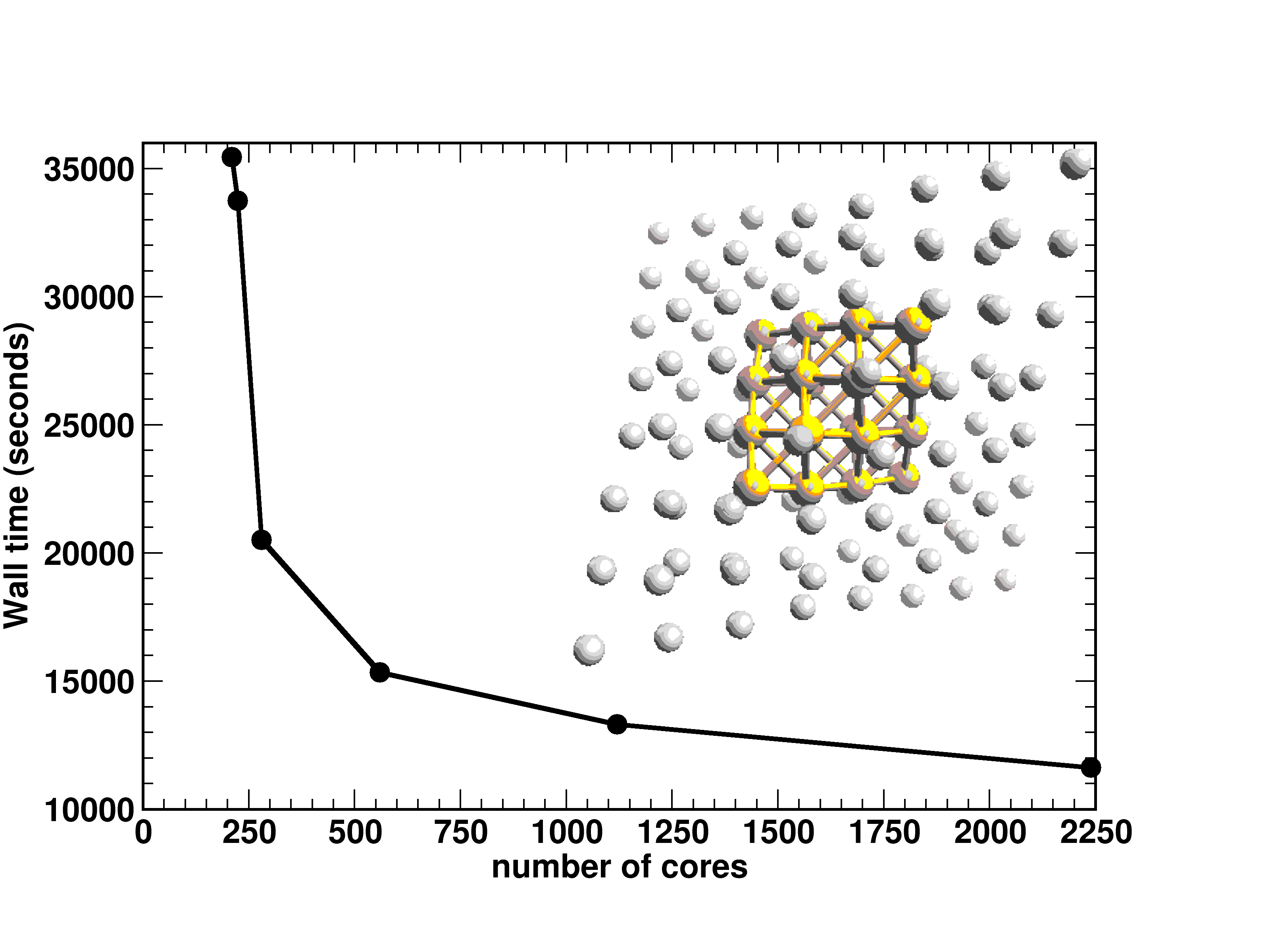}
\caption{LR-TDDFT benchmark for the Au$_{20}$ molecule in a neon matrix.}
\label{fig:tddft_au20}
\end{figure}

In Fig.~\ref{fig:tddft_au20}, we report the parallel performance of the Gaussian basis set LR-TDDFT module in NWChem.
This calculation involved computing 100 excitation energies,
requiring 11 Davidson iterations, for
the Au$_{20}$ molecule surrounded by a matrix of 80 Ne atoms\cite{yu2017}
(1840 basis functions) with $D_2$ symmetry, using the B3LYP functional.
These calculations were performed on the Cascade
supercomputer located at PNNL.\\

\fi

\paragraph{Closed-shell CCSD(T):}
The parallel implementation of the CCSD(T) approach by  Kobayashi and Rendell \cite{Kobayashi}, employing the spin adaptation scheme based on the unitary group approach (UGA) 
\cite{Scuseria} within NWChem, was one of the  first scalable  implementations of the 
CC formalism  capable of taking advantage of several hundred processors.
This implementation was  used in 
simulations involving tera- and peta-scale architectures where chemical accuracy is required to
describe ground-state potential energy surfaces. One of the best illustrations of the performance of the CCSD(T) implementation is provided by
 calculations for water clusters \cite{apra2009liquid}. In  the largest calculation,  (H$_2$O)$_{24}$, sustained performance of 1.39 
PetaFLOP/s (double precision) on 223,200 processors of ORNL's Jaguar system  was documented. 
This impressive performance was mostly 
attributed to the (T)-part characterized by $n_o^3n_u^4$ numerical overhead (where $n_o$ and $n_u$ refer to the total numbers of correlated occupied and virtual orbitals)  and its relatively low communication footprint. \\

\begin{figure}[t]
\centering
\includegraphics[scale=0.32]{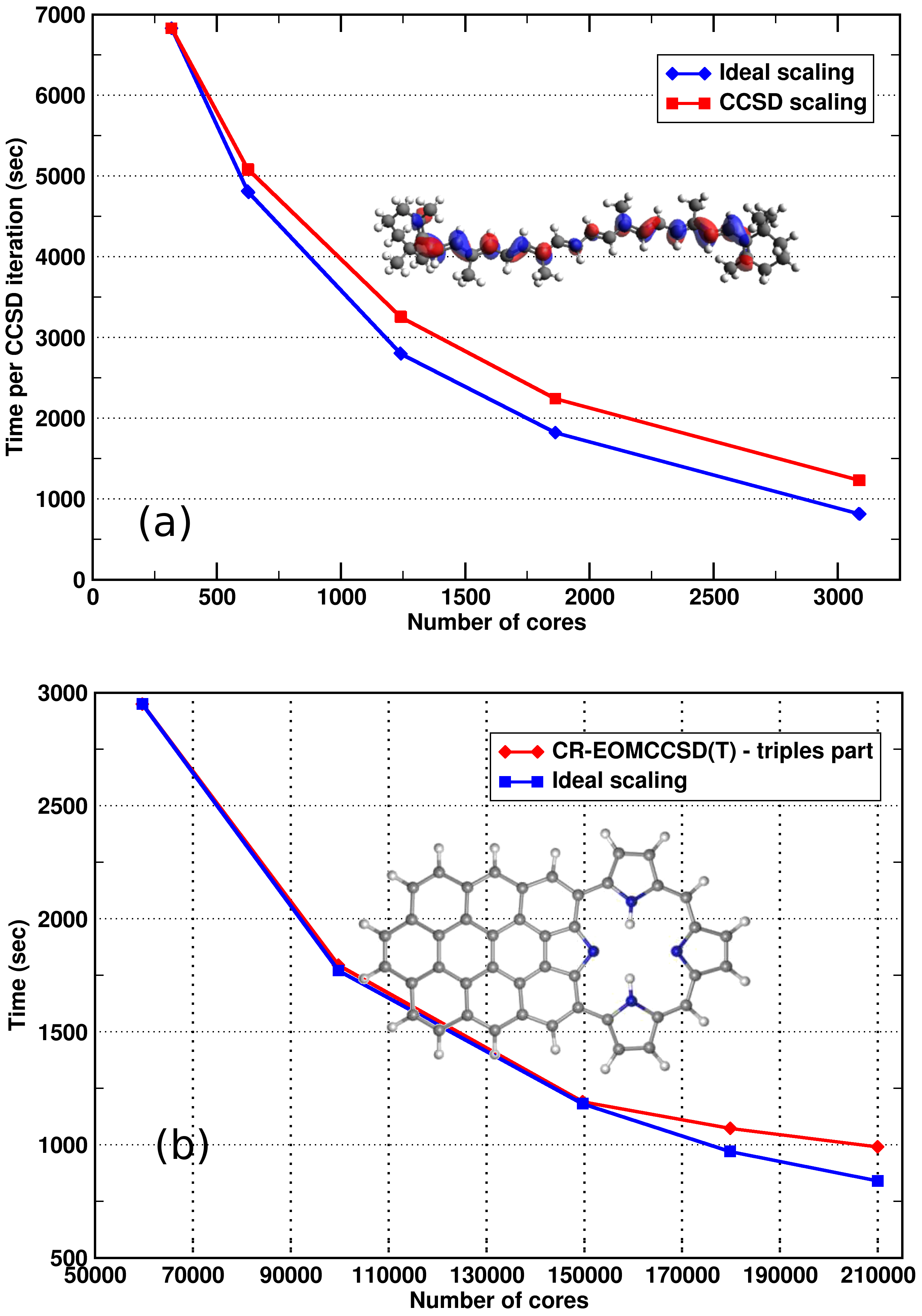}
\caption{
Benchmark EOMCC scalability tests for : (a) $beta$-carotene and (b) free-base porphyrin (FBP) fused coronene.
Timings for CR-EOMCCSD(T) approach for the coronene fused free-base porphyrin in the AVTZ basis set were determined from calculations on the ORNL's Jaguar Cray XT5 computer system.
}
\label{creom}
\end{figure}

\paragraph{Tensor Contraction Engine:}
The TCE has enabled parallel CC/EOMCC/LR-CC  calculations for closed- and open-shell systems characterized by 1,000-1,300 orbitals.
Some of the most illustrative examples include calculations for static and frequency-dependent polarizabilities for polyacenes and  $C_{60}$ molecule,\cite{hammond2007dynamic,kowalski2008coupled}
excited state simulations for $\pi$-conjugated chromophores,\cite{doi:10.1021/ct200217y} and IP-EOMCCSD calculations for carbon nanotubes.
\cite{peng2017coupled}
A good illustration of the scalability of the TCE module is provided by the  application of  GA-based TCE implementations of the iterative (CCSD/EOMCCSD) and non-iterative (CR-EOMCCSD(T)) methods  in studies of excited states of 
$\beta$-carotene \cite{hu2014toward} and 
functionalized forms of porphyrin \cite{kowalski2011scalable}
(see Fig.\ref{creom}(a) and (b), respectively). While non-iterative methods are much easier to scale across a
large number of cores (Fig.\ref{creom} (b)),  scalability of the iterative CC methods is less easy to achieve. 
However, using early task-flow algorithms for TCE CCSD/EOMCCSD methods 
\cite{kowalski2011scalable}
it was possible to achieve satisfactory scalability  in the range of 1,000-8,000 cores.\\

\paragraph{Recent Implementation of Plane-Wave DFT AIMD for Many-Core Architectures:} 
\label{pwperformance}
The very high degree of parallelism available on machines with many-core processors is forcing developers to carefully revisit the implementation of their programs in order to make use of this hardware efficiently. In this section, after a brief overview of the computational costs and parallel strategies for AIMD, we present our recent work~\cite{bylaskaintel17} on adding thread-level parallelism to the AIMD method implemented in NWChem~\cite{VALIEV20101477,de2010utilizing,bylaska2011large}.

\begin{figure}[ht] 
\centering
\includegraphics[scale=0.28]{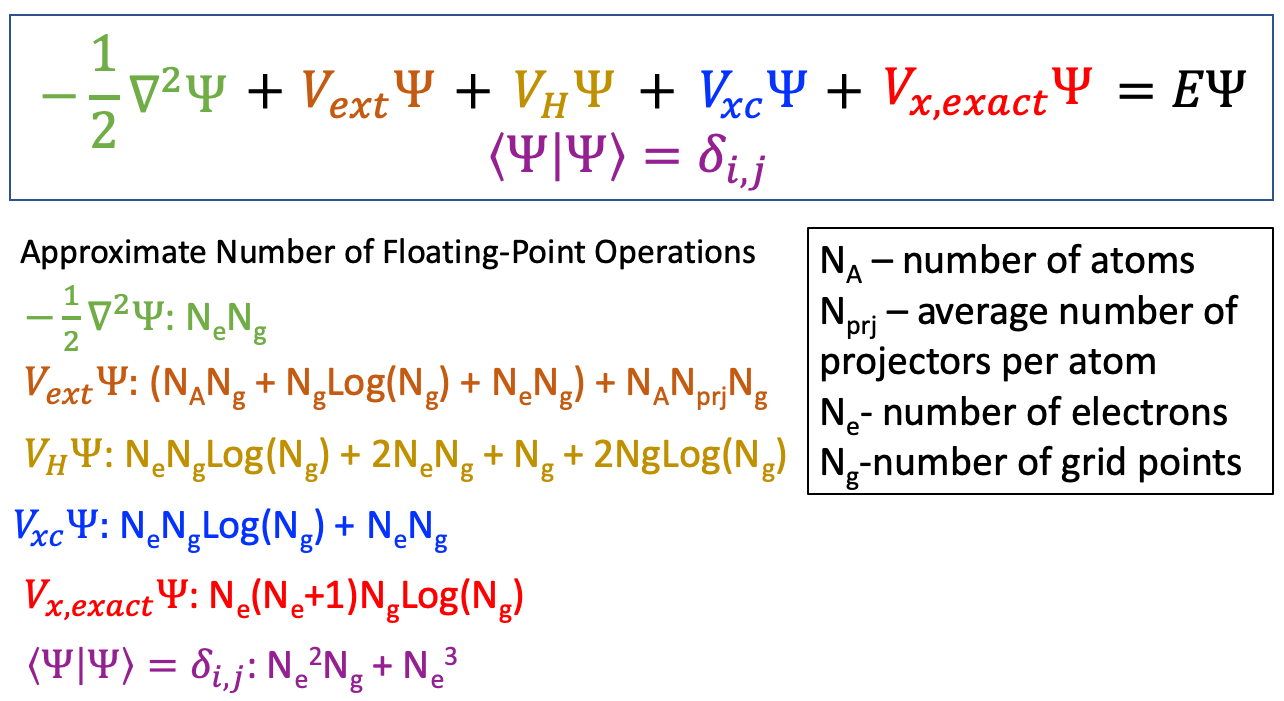}
\caption{Operation count of $H\psi_i$ in a plane-wave DFT simulation. Figure from Ref. \onlinecite{Mathias17}}
\label{nwpwops}
\end{figure}

The main computational costs of an energy minimization or AIMD simulation are the evaluation of the electronic gradient
$\delta E_{total} / \delta \psi_i^{*} = H\psi_i$ and algorithms used to maintain orthogonality.
These costs are illustrated in Figure ~\ref{nwpwops}. 
Due to their computational complexity, the electron gradient $H \psi_i$ and orthogonalization need to be calculated as efficiently as possible.
The main parameters that determine the cost of a calculation are $N_g$, $N_e$, $N_a$, and $N_{proj}$, where $N_g$ is the size of the three-dimensional FFT grid, $N_e$ is the number of occupied orbitals, $N_a$ is the number of atoms, $N_{proj}$ is the number of projectors per atom, and $N_{pack}$ is the size of the reciprocal space. 

The evaluation of the electron gradient (and orthogonality) contains three major computational pieces that need to be efficiently parallelized:
\begin{itemize}
\setlength{\itemsep}{0pt}%
\setlength{\parskip}{0pt}
\item \emph{applying} $V_H$ and $V_{xc}$, involving the calculation of $2N_e$ 3D FFTs;
\item \emph{calculating the non-local pseudopotential}, $V_{NL}$, dominated by the cost of the matrix multiplications $W = P^{T}Y$, and $Y_2 =PW$, where $P$ is an {$N_{pack}\times (N_{proj} \cdot N_a)$} matrix, $Y$ and $Y_2$ are
  {$N_{pack} \times N_e$} matrices, and $W$ is an {$(N_{proj}N_a) \times N_e$} matrix;
\item \emph{enforcing orthogonality}, where the most expensive matrix multiplications are $S=Y^{T} Y$ and $Y_2 = Y S$, where $Y$ and $Y_2$ are {$N_{pack} \times N_e$} matrices, and $S$ is an {$N_e \times N_e$} matrix.  In this work, Lagrange multiplier kernels are used for maintaining orthogonality of Kohn-Sham orbitals\cite{nelson1993plane,wiggs1995hybrid,bylaska2011large,canning2005scaling,gygi2008architecture}.
\end{itemize}

\begin{figure}[t]  
     \centering
     \begin{adjustbox}{width=\linewidth}
     \input{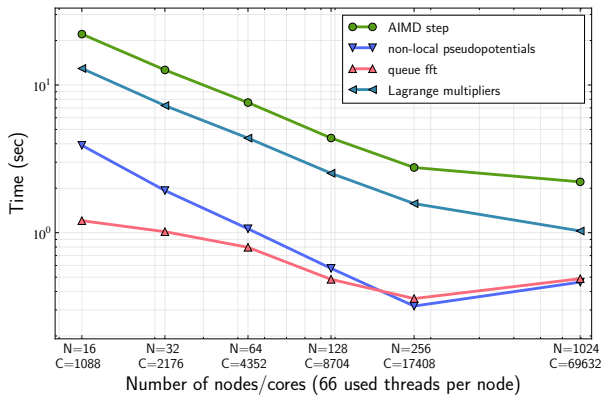}
     \end{adjustbox}
     \caption{Scalability of major components of an AIMD step on the Xeon Phi partition for a simulation
 of    256 H$_2$O   molecules.  Figure from Bylaska {\it et al}.~\cite{bylaskaintel17}.}
     \label{fig.ss.knl.256}
\end{figure}

In Fig.~\ref{fig.ss.knl.256} the timing results for a full AIMD
simulation of 256 water molecules on 16, 32, 64, 128, 256, and 1024
KNL nodes are shown.  The ``Cori'' system at NERSC was used to run
this benchmark. This benchmark was taken from Car Parrinello
simulations of 256 H$_2$O with an FFT grid of $N_g=180^3$ ($N_e$=2056)
using the plane-wave DFT module (PSPW) in NWChem. In these timings,
the number of threads per node was 66. The size of this benchmark
simulation is about 4 times larger than many mid-size AIMD simulations
carried out in recent years, e.g. in recent work by
Bylaska and co-workers.~\cite{swaddle2005kinetic,rustad2007ab,atta2012structure,fulton2012near,solve1,solve2}
The overall timings show strong scaling up to 1024 KNL nodes (69632
cores) and the timings of the major kernels, the pipelined 3D FFTs,
non-local pseudopotential, and Lagrange multiplier kernels all
displayed significant speedups.\\

\paragraph{Classical Molecular Dynamics:} 
The molecular dynamics module in the current NWChem release is based
on the distribution of cells over available ranks in the
calculation. Simulations exhibit good scalability when cells only
require communication with immediately neighboring cells. When the
combination of cell size and cutoff radius is such that interactions
with atoms in cells beyond the immediate neighbors are required,
performance is significantly affected. This limits the number of ranks
that can effectively be used. For example, a system with 500,000 atoms
will only scale well up to 1000 ranks. In future implementations, the cell-cell
pair-list will be distributed over the available ranks. While this
leads to additional communication for ranks that do not ``own'' a cell,
the implementation of a new communication scheme that avoids global
communication has been demonstrated to improve scalability by at least
an order of magnitude.\cite{Straatsma2013}

\section{Outreach}
\begin{figure}[t]
\centering
\includegraphics[scale=0.27]{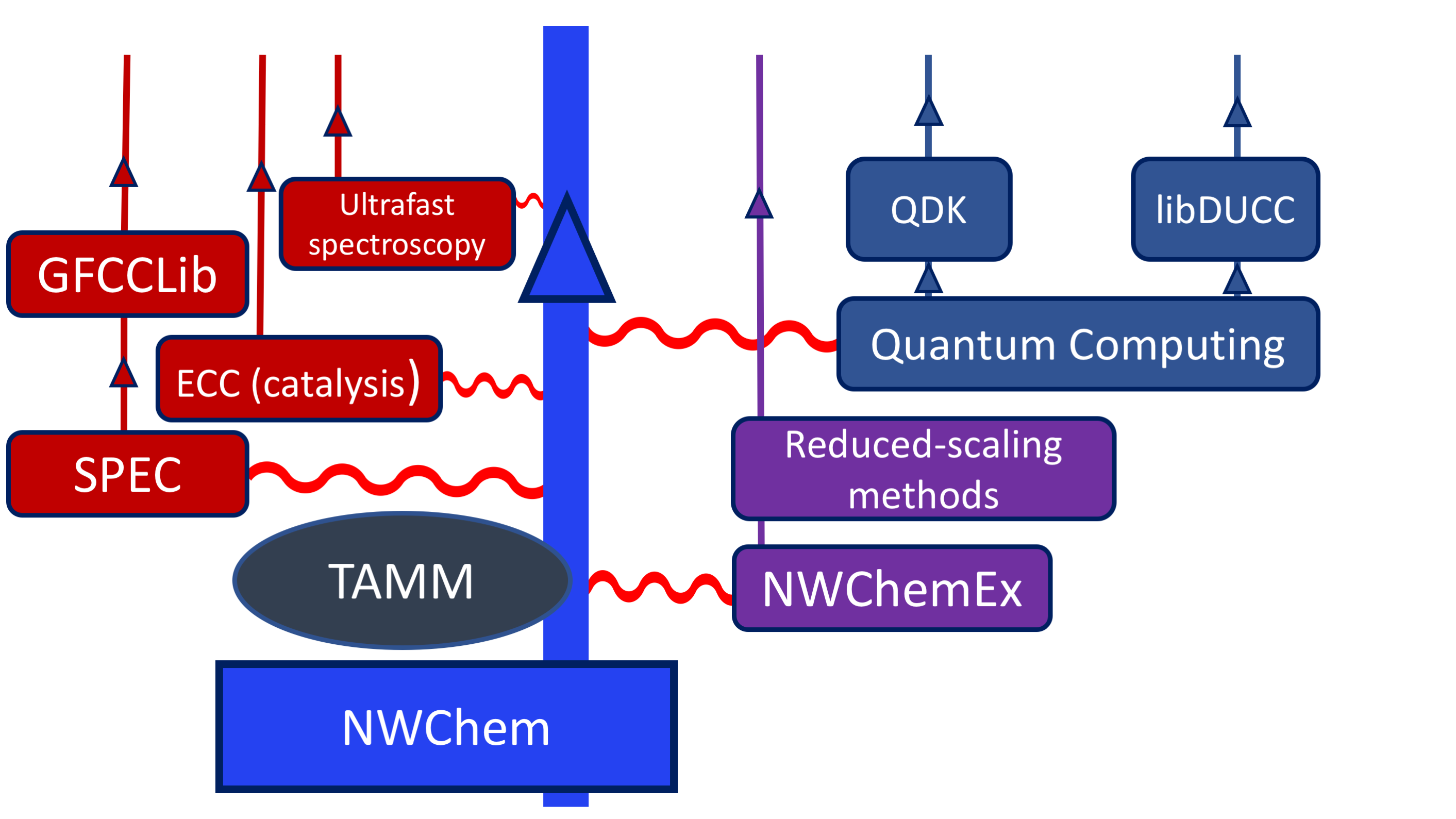}
\caption{A ``connected diagram'' describing ongoing efforts to 
extend computational chemistry models to exa-scale  and quantum computing. In each case, NWChem provides a testing and development platform. A significant role in these projects is played by  Tensor Algebra for Many-body Methods (TAMM) library. The ECC acronym stands for the Exa-scale Catalytic Chemistry project supported by BES.
\cite{mutlu2019toward}
The QDK-NWChem interface with the libDUCC library is used for
downfolding electronic Hamiltonians.\cite{bauman2019downfolding} 
}
\label{nwchem_future}
\end{figure}

%
Given the various electronic structure methods available in NWChem, 
it does not come as a surprise that many of these functionalities 
have been integral to various projects focused on extensions of  quantum chemical capabilities to exa-scale architectures and emerging quantum computing
(see Fig.~\ref{nwchem_future}
for a pictorial representation of recent developments).
Below we describe several examples of such a synergy.\\


\paragraph{Interfacing with Other Software:}
Over the years, many open-source and commercial developers have been using NWChem as a resource for their capability development, and building add-on tools to increase the code's usability. 
Various open-source and commercial platforms provide user interface capabilities to set up and analyze the results of calculations that can be performed with NWChem.~\cite{hanwell2012, Hjorth_Larsen_2017,jmol,chemcraft,mocalc2012, gausssum,Chemissian,scienomics,Ascalaph,emslarrows} NWChem initially developed its own graphical user interface called the Extensible Computational Chemistry Environment~\cite{ecce}, which is currently supported by a group of open-source developers.
In addition, multiple codes use quantities from the NWChem simulation, such as wavefunctions as input for the calculation of additional properties not directly available in the code.~\cite{CamCASP,chemshell,NIKOLAIENKO201415,doi:10.1002/jcc.23470,LOURDERAJ20141074, fiesta,pupil,votca,pydp4,fafoom}
NWChem is able to export electrostatic potential and charge densities
with the Gaussian cube format~\cite{gaussiancube}
and can use the Molden format~\cite{Schaftenaar2017}  to write or read
molecular orbitals.
This allows codes~\cite{molekel,BERGMAN1997301,HUMP96,jamberoo,Momma:ko5060} to utilize NWChem's data to, for example, display charge densities and electrostatic potentials. NWChem can also generate AIM wavefunction files that have been used by a variety of codes to calculate various properties.~\cite{doi:10.1021/ct100641a,XAIM,xdm,doi:10.1002/jcc.22885} Recently, NWChem has also been interfaced with the SEMIEMP code\cite{SEMIEMP}, which can be used to perform real-time electronic dynamics using the INDO/S Hamiltonian.\cite{ghosh2017,ghosh2019} \\
 %

\paragraph{Common Component Architecture:} It is an attractive idea to encapsulate complex scientific
applications as components with standardized interfaces. The
components interact only through these well-defined interfaces and can
be combined into full applications. The main motivation is to be able
to reuse and swap components as needed and seamlessly create complex
applications. There have been a few attempts to introduce this
approach to the scientific software development community. The most
notable DOE-led effort was the Common Component Architecture (CCA)
Forum\cite{cca1}, which was launched in 1998 as a scientific community effort
to create components designed specifically for the needs of
high-performance scientific computing. A more recent development is
the rise of Simulation Development Environment (SDE) framework\cite{cca2},
which has features that are related to the components of CCA.

NWChem developers have participated in CCA and SDE effort resulting in the
creation of the NWChem component. As an example, the NWChem CCA component was
used in the building applications for molecular geometry optimization
from multiple quantum chemistry and numerical optimization packages
\cite{cca3}, combination of multiple theoretical methods to improve
multi-level parallelism \cite{cca4}, demonstration of multi-level parallelism
\cite{cca5}, and standardization of integral interfaces in quantum chemistry
\cite{cca6}.
In the end, the CCA framework was too cumbersome to use for
developers, requiring significant efforts to develop interfaces and
making components to work together. It resulted in the retirement of
CCA Forum in 2010, but the work done on standardization of interfaces
is continuing to benefit the quantum chemistry community to this day.\\

\paragraph{NWChemEx:} The NWChemEx project is a natural extension of NWChem to
overcome the scalability challenges associates with migrating  the current code base to  exa-scale platforms. 
NWChemEx is being developed to address two outstanding problems 
in advanced biofuels research: 
(i ) development of a molecular understanding of proton controlled membrane transport processes  and (ii) development of catalysts for the efficient conversion of biomass-derived
intermediates into biofuels, hydrogen, and other bioproducts; 
therefore the main focus is on enabling scalable implementations of the ground-state canonical CC formalisms utilizing Cholesky decomposed form of the two-electron integrals 
\cite{bostrom2012coupled,pedersen2004polarizability,epifanovsky2013general,feng2019,peng2019coupled,folkestad2019efficient}
as well as linear scaling CC formulations based on the 
domain-based local pair natural orbital CC formulations (DLPNO-CC).\cite{riplinger2013efficient,ripliner2016,pavosevic2017} and embedding methods.\\

\paragraph{Scalable Predictive methods for  Excitations and Correlated phenomena (SPEC):} The main focus of the SPEC  software project 
is to provide the users with a new generation of methodologies 
to simulate excited states and excited-state processes using existing peta- and emerging exa-scale architectures.  
These new capabilities will play an important role in supporting the experimental efforts at light source facilities, which require accurate and reliable modeling tools. The existing NWChem capabilities are being used to verify and validate SPEC implementations including excitation energy, ionization potential, 
and electron affinity variants of the EOMCC theory as well as hierarchical Green's function formulations ranging from the lower order GW$+$Bethe-Salpeter equation (GW$+$BSE)\cite{onida2002} to hierarchical coupled-cluster Green's function (GFCC) methods \cite{nooijen92_55,nooijen93_15,nooijen95_1681,meissner93_67,peng2018green} and multi-reference CC methods. \\

\paragraph{Quantum Information Sciences:} Quantum computing not only offers the promise of  
overcoming exponential computational barriers of conventional computing but also in achieving 
the ultimate level of accuracy in studies of challenging processes involving 
multi-configurational states in  catalysis, biochemistry, photochemistry, and materials science to name only a few areas where quantum information technologies can lead to the transformative changes in the way how quantum simulations are performed. NWChem,  with its computational infrastructure to characterize second-quantized forms of electronic Hamiltonians in various basis sets (Gaussian and plane-waves) and 
with wavefunction methodologies to provide an initial characterization of the ground- and excited-state wavefunctions,  
can be used as a support platform for various types of quantum simulators. 
The recently developed QDK-NWChem interface \cite{low2019q} (QDK designates Quantum Development Kit developed by Microsoft Research team) 
for quantum simulations and libraries for CC downfolded electronic Hamiltonians for quantum computing  \cite{bauman2019downfolding}
are good illustrations of the utilization of NWChem in supporting the quantum computing effort.

\section{Data Availability Statement}
The data that support the findings of this study are available from the corresponding author upon reasonable request.

\section{Conclusions}
The NWChem project is  an example of a successful co-design effort that harnesses 
the expertise and experience  of researchers in several 
complementary areas, including quantum chemistry, applied mathematics, and high-performance computing.
Over the last three decades, NWChem has evolved into a code that offers a unique combination of computational tools to tackle complex 
chemical processes at various spatial and time scales.  

In addition to the development of new methodologies, NWChem is being continuously upgraded, with new algorithms, to take advantage of emerging computer architectures and quantum information technologies. We believe the community model of NWChem will continue to spur exciting new developments well into the future.

\section*{Supplementary Material}
See supplementary material for tutorial slides showing examples of NWChem input files.

\section*{Acknowledgments}
The authors wish to acknowledge the important contributions to the NWChem project of the following deceased researchers:
Ricky A. Kendall, Jarek Nieplocha and Daniel W. Silverstein.

We would like to thank people that have helped the progress of the
NWChem project by providing valuable feedback, scientific direction,
suggestions, discussions and encouragement. This (incomplete) list includes:
A. Andersen, J. Andzelm, R. Baer, A. Bick, V. Blum, D. Bowman, E.~A. Carter,
A. Chakraborty, G. Cisneros, D. Clerc, A.~J. Cohen,  L.~R Corrales, A.~R. Felmy, A. Fortunelli,
J. Fulton, D.~A. Dixon, P.~Z. El-Khoury, D. Elwood, G. Fanourgakis, D. Feller,
B.~C. Garrett, B. Ginovska, E. Glendening, V.~A. Glezakou, M.~F. Guest, M. Gutowski, M. Hackler, M. Hanwell, 
W.~L. Hase, A.~C. Hess, W.~M. Holden, C. Huang, W. Huang, E.~S. Ilton, E.~P. Jahrman, J. Jakowski, J.~E. Jaffe, J. Ju,
S.~M. Kathmann, R. Kawai, M. Khalil, X. Krokidis, L. Kronik, P.~S. Krsti\'{c}, R. Kutteh,
X. Long, A. Migliore, E. Miliordos, M. Nooijen, J. Li, L. Lin, S. Liu, N. Maitra, M. Malagoli, B. Moore
II, P. Mori, S. Mukamel, C. J. Mundy, V. Meunier, V. Murugesan, D. Neuhauser,
S. Niu, R.~M. Olson, S. Pamidighantam, M. Pavanello, M.~P. Prange, C.~D. Pemmaraju, D. Prendergast, S. Raugei, J.~J. Rehr, 
A.~P. Rendell, R. Renslow, T. Risthaus, K.~M. Rosso, R. Rousseau, 
J.~R. Rustad,  
G. Sandrone, G. K. Schenter, G.~T. Seidler, H. Sekino, P. Sherwood, C. Skylaris, H. Song, L. Subramanian, B.~G. Sumpter,
P. Sushko, S. Tretiak, R.~M. Van Ginhoven, B.~E. Van Kuiken, J.~H. van Lenthe, B. Veeraraghavan,
E. Vorpagel, X.~B. Wang, J. Warneke, Y. A. Wang, J.~C. Wells, S.~S. Xantheas,
W. Yang, J. Zador, Y. Zhang.

The core development team acknowledges support from the following projects at the Pacific Northwest National
Laboratory. Pacific Northwest National Laboratory is operated by
Battelle Memorial Institute for the U.S. Department of Energy under Contract DE-AC05-76RL01830: 
(i) Environmental and Molecular Sciences Laboratory (EMSL), the Construction Project, and Operations, the Office of Biological and Environmental Research, 
(ii) the Office of Basic Energy Sciences, Mathematical, Information, and Computational Sciences, Division of Chemical Sciences, Geosciences, and Biosciences (CPIMS, AMOS, Geosciences, Heavy Element Chemistry, BES Initiatives:~CCS-SPEC, CCS-ECC, BES-QIS, BES-Ultrafast), 
(iii) the Office of Advanced Scientific Computing Research through the Scientific Discovery through Advanced Computing (SciDAC), Exascale Computing Project (ECP): NWChemEx.
Additional funding was provided by the Office of Naval Research, the U.S. DOE High Performance Computing and Communications Initiative and industrial collaborations (Cray, Intel, Samsung).

The work related to the development of QDK-NWChem interface was
supported by the ``Embedding Quantum Computing into Many-body
Frameworks for Strongly Correlated Molecular and Materials Systems''
project, which is funded by the U.S. Department of Energy(DOE), Office
of Science, Office of Basic Energy Sciences, the Division of Chemical
Sciences, Geosciences, and Biosciences, and Quantum Algorithms,
Software, and Architectures (QUASAR) Initiative at Pacific Northwest
National Laboratory (PNNL).

S. A. Fischer acknowledges support from the U.S. Office of Naval Research
through the U.S. Naval Research Laboratory.

A. J. Logsdail acknowledges support
from the UK EPSRC under the ``Scalable Quantum Chemistry with Flexible
Embedding'' Grants EP/I030662/1 and EP/K038419/1. 

A. Otero-de-la-Roza  acknowledges support from the Spanish government for a
Ram\'on y Cajal fellowship (RyC-2016-20301) and for financial support
(projects PGC2018-097520-A-100 and RED2018-102612-T).

J. Autschbach acknowledges support from the U.S. Department of Energy,
Office of Basic Energy Sciences, Heavy Element Chemistry program
(Grant No. DE-SC0001136, relativistic methods \& magnetic resonance
parameters) and the National Science Foundation (Grant
No. CHE-1855470, dynamic response methods).

D. Mejia Rodriguez acknowledges support from the U.S. Department of Energy Grant No. DE-SC0002139.

D.~G. Truhlar acknowledges support from the NSF under grant no. CHE–1746186.

E.~D. Hermes was supported by the U.S. Department of Energy,
Office of Science, Basic Energy Sciences, Chemical Sciences,
Geosciences and Biosciences Division, as part of the
Computational Chemistry Sciences Program (Award Number:
0000232253). Sandia National Laboratories is a multimission
laboratory managed and operated by National Technology and
Engineering Solutions of Sandia, LLC, a wholly owned
subsidiary of Honeywell International, Inc., for the U.S.
Department of Energy’s National Nuclear Security Administration
under contract DE-NA0003525.

Z. Lin acknowledges support from the National Natural Science Foundation of China (Grants 11574284 and 11774324).

J. Garza thanks CONACYT for the project FC-2016/2412.

K. Lopata gratefully acknowledges support by the U.S. Department of Energy, Office of Science, Basic Energy Sciences, Early Career Program, under Award No. DE-SC0017868 and U.S. Department of Energy, Office of Science, Basic Energy Sciences, under Award No. DE-SC0012462

L. Gagliardi and C.~J. Cramer acknowledge support from by the
Inorganometallic Catalyst Design Center, an Energy Frontier Research
Center funded by the US Department of Energy (DOE), Office of Science,
Basic Energy Sciences (BES) (DE-SC0012702). They acknowledge the
Minnesota Supercomputing Institute (MSI) at the University of
Minnesota for providing computational resources.

This work used resources provided by EMSL, a DOE Office of Science User Facility sponsored by the Office of Biological and Environmental Research and located at the Pacific Northwest National Laboratory (PNNL) and PNNL Institutional Computing (PIC). PNNL is operated by Battelle Memorial Institute for the United States Department of Energy under DOE Contract No. DE-AC05-76RL1830. 

The work also used resources provided by the National Energy Research Scientific Computing Center (NERSC), a DOE Office of Science User Facility supported by the Office of Science of the U.S. Department of Energy under Contract No. DE-AC02-05CH11231.

This work also used resources provided by Oak Ridge Leadership Computing Facility (OLCF) at the Oak Ridge National Laboratory, which is
supported by the Office of Science of the U.S. Department of Energy
under Contract No. DE-AC05-00OR22725.

This manuscript has
been authored in part by UT-Battelle, LLC, under contract
DE-AC05-00OR22725 with the US Department of Energy (DOE). The US
government retains and the publisher, by accepting the article for
publication, acknowledges that the US government retains a
nonexclusive, paid-up, irrevocable, worldwide license to publish or
reproduce the published form of this manuscript, or allow others to do
so, for US government purposes. DOE will provide public access to
these results of federally sponsored research in accordance with the
DOE Public Access Plan
(https://energy.gov/downloads/doe-public-access-plan).

\iftrue
This article may be downloaded for personal use only. Any other use
requires prior permission of the author and AIP Publishing. This
article appeared in volume 152, issue 18, page 184102 of the Journal of Chemical Physics and may be 
\fi
found at
\url{https://doi.org/10.1063/5.0004997}

\bibliography{references}

\end{document}